%% file: flow-algorithms.tex
\documentclass{article}
\input{preamble-graphth-charn.tex}

\title{A complete algorithm to find flows in the one-way measurement model}
\author{Niel de Beaudrap~\footnote{This work was supported in part by ARDA, ORDCF, MITACS,
and CIAR.}\\ IQC, University of Waterloo \\ \texttt{jdebeaud@iqc.ca} }
\date{}

\begin{document}
\maketitle

\begin{abstract}
	This article is the complement to~\cite{B06b}, which proves that flows (as introduced by~\cite{DK05})
	can be found efficiently for patterns in the one-way measurement model which have non-empty
	input and output subsystems of the same size. This article presents a complete algorithm for
	finding flows, and a proof of its' correctness, without assuming any knowledge of graph-theoretic
	algorithms on the part of the reader.	This article is a revised version of~\cite{B06a}, where the
	results of~\cite{B06b} also first appeared.
	\vspace{-2.05ex}
\end{abstract}

\section{Introduction}

In the one-way measurement model~\cite{RB01,RB02,RBB03}, algorithms are essentially described by
a sequence of single-qubit measurements (where the choice of measurement may depend on earlier
measurement results in a straightforward way) performed on a many-qubit entangled state. This
many-qubit state may be described in terms of the state of an \emph{input system} $I$\,, together
with a graph $G$ of \emph{entangling operations} involving $I$ and a collection of auxiliary
qubits prepared in the $\ket{+}$ state: each edge of $G$ represents a single controlled-$Z$
operation between two qubits. After the sequence of measurements, any qubits left unmeasured
still support a quantum state, and are interpreted as an \emph{output system} $O$\,. A triple
$(G,I,O)$ belonging to a given pattern is called the \emph{geometry} of the pattern.

In~\cite{B06a}, it was shown that the flow property defined by Danos and Kashefi~\cite{DK05} can
be efficiently tested for a geometry $(G,I,O)$ when $\abs{I} = \abs{O}$\,. The property is the existence
of a \emph{causal flow}\footnote{These are simply called ``flows'' in~\cite{DK05}: I use the term
``causal flow'' in this article to maintain consistency with~\cite{B06b}.}, which describes a partial
order $\preceq$ describing an order (independent of measurement angles) in which the qubits of the
geometry may be measured to perform a unitary embedding, once suitable corrections are applied to
the output qubits. Causal flows may allow quantum algorithms to be devised in the one-way measurement model
without using the circuit model:~\cite{BDK06} proposes one way in which this might be done.

This article presents a complete algorithm for finding causal flows for a geometry $(G,I,O)$
with $\abs{I} = \abs{O}$ in time $O(km)$\,, where $k = \abs{I} = \abs{O}$ and $m = \abs{E(G)}$\,, 
suitable for an audience with no experience in graph-theoretic algorithms. This is a revised
version of~\cite{B06a}, re-written with the aim of focusing on the algorithm for finding flows
for the sake of reference. For the graph-theoretical characterization of flows, this article
refers to~\cite{B06b}, which is an improved presentation of the graph-theoretic results
presented originally in~\cite{B06a}.

Although no knowledge of graph-theoretic algorithms is assumed, a basic understanding of graph theory and the
one-way measurement model is essential. For basic definitions in graph theory, readers 
may refer to Diestel's excellent text~\cite{Diestel}; I will use the conventions of~\cite{B06b, DKP04b}
for describing patterns in the one-way model.

\section{Preliminaries}
\label{sec:preliminaries}

In this section, we will fix our conventions and review the results and terminology
of~\cite{B06b}.

\subsection{Basic notation and conventions}

For a graph $G$\,, we write $V(G)$ for the set of vertices and $E(G)$ for the set
of edges of $G$\,. Similarly, for a directed graph (or \emph{digraph}) $D$\,, we write 
$V(D)$ for the set of vertices and $A(D)$ for the set of directed edges (or \emph{arcs})
of $D$\,. If $x$ and $y$ are adjacent, we let $xy$ denote the edge between them in a graph,
and $x \arc y$ denote an arc from $x$ to $y$ in a digraph. We will use the convention that
digraphs may contain loops on a single vertex and multiple edges between two vertices, but
that graphs cannot have either.

When a graph $G$ is clear from context, we will write $x \sim y$ when $x$ and $y$ are adjacent
in $G$\,,  and write $S\comp$ to represent the complement of a set of vertices $S \subset V(G)$\,.
 
If $\mC$ is a collection of directed paths (or \emph{dipaths}), we will say that $x \arc y$
is an \emph{arc of $\mC$}, and that the edge $xy$ is \emph{covered} by $\mC$\,, when
$x \arc y$ is an arc in a path $P \in \mC$\,. 

In this paper, $\N$ denotes the non-negative integers. For any $n \in \N$\,,
$[n]$ denotes the set $\set{j \in \N}{j < n}$\,.

\subsection{Results for Causal Flows}

\subsubsection{Definition and motivation}

\begin{definition}
	\label{dfn:causalFlow}
	A \emph{geometry} $(G, I, O)$ is a graph $G$ together with subsets $I, O \subset V(G)$\,.
	We call $I$ the \emph{input vertices} and $O$ the \emph{output vertices} of the geometry.
	A \emph{causal flow} on $(G,I,O)$ is an ordered pair $(f, \preceq)$\,, with a function
	$f: O\comp \to I\comp$ and a partial order $\preceq$ on $V(G)$\,, such that
	\begin{align}
		\flowi\quad
	&
		x \sim f(x)\,;
	&
		\flowii\quad
	&
		x \preceq f(x)\,;
	&
		\flowiii\quad
	&
		y \sim f(x) \;\implies\; x \preceq y\,,
	\end{align}
	hold for all vertices $x \in O\comp$ and $y \in V(G)$\,. We will refer to $f$ as the
	\emph{successor function} of the causal flow, and $\preceq$ as the \emph{causal
	order} of the causal flow.
\end{definition}

A geometry $(G,I,O)$ represents the information of a one-way measurement pattern which
is independent of the order of operations and measurement angles. $G$ is the entanglement
graph of the pattern, $I$ is the set of qubits which are not prepared in a fixed state
initially (their joint initial state in the algorithm may be arbitrary), and $O$
represents the set of qubits which are not measured in the pattern (which thus support
a final quantum state).

The conditions~\flowi~--~\flowiii\ are motivated by how byproduct operators and signal dependencies
are induced by commuting correction operations to the end of a pattern which performs a unitary embedding.
The significance of a causal flow on a geometry $(G,I,O)$ is that any pattern defined on that geometry
can be transformed into one which has the same measurement angles and which performs a unitary embedding
$\mathcal H_I \to \mathcal H_O$\,. In particular, this means that unitary embeddings can be devised
in the measurement model by ignoring signal dependencies and treating each measurement operator as though
it post-selects for some one of the states in the basis of the measurement.
See Section~2.2 of~\cite{B06b} for details.

\subsubsection{Graph-theoretic characterization}

The result of~\cite{B06b} was obtained by characterizing causal flows in terms of collections of
vertex-disjoint paths.
\begin{definition}
	\label{dfn:pathCoverSuccFn}
	Let $(G,I,O)$ be a geometry. A collection $\mC$ of (possibly trivial) directed paths in
	$G$ is a \emph{path cover} of $(G,I,O)$ if
	\begin{romanum}
	\item
		each $v \in V(G)$ is contained in exactly one path (i.e. the paths
		cover $G$ and are vertex-disjoint);
	\item
		each path in $\mC$ is either disjoint from $I$\,, or intersects $I$ only at its
		initial point;
	\item
		each path in $\mC$ intersects $O$ only at its final point.
	\end{romanum}
	The \emph{successor function} of a path cover $\mC$ is the unique $f: O\comp \to I\comp$ such
	that $y = f(x)$ if and only if $x \arc y$ is an arc of $\mC$\,. If a function
	$f: O\comp \to I\comp$ is a successor function of \emph{some} path-cover of $(G,I,O)$\,,
	we call $f$ a \emph{successor function of $(G,I,O)$}.
\end{definition}

If a geometry $(G,I,O)$ has a causal flow $(f,\preceq)$\,, the maximal orbits of the successor function
$f$ define a path cover for $(G,I,O)$\,, which allow us to consider the causal flow in terms of
vertex-disjoint paths in $G$\,:

\begin{citetheorem}[Lemma~3]{B06b}{Theorem}
	\label{thm:orbitPathCover}
	Let $(f, \preceq)$ be a causal flow on a geometry $(G,I,O)$\,. Then there is a path cover
	$\mP_f$ of $(G,I,O)$ whose successor function is $f$\,.
\end{citetheorem}

\begin{figure}[ht]
	\setlength\unitlength{0.16mm}
	\begin{center}
 		\begin{picture}(500,175)(-20,0)
 			\includegraphics[width=24mm]{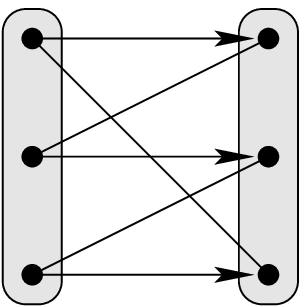}
 			\puttext(-140,-10)[c]{$I$}
 			\puttext(-20,-10)[c]{$O$}
 			\puttext(-155,140)[r]{$a_0$}
 			\puttext(-155,80)[r]{$a_1$}
 			\puttext(-155,20)[r]{$a_2$}
 			\puttext(25,140)[r]{$b_0$}
 			\puttext(25,80)[r]{$b_1$}
 			\puttext(25,20)[r]{$b_2$}
			\puttext(190,110)[c]{$\flowii \;\implies\;
												\begin{cases}
														a_0 &\!\!\!\!\preceq b_0	\\
														a_1 &\!\!\!\!\preceq b_1	\\
														a_2 &\!\!\!\!\preceq b_2
												\end{cases}$}
			\puttext(190,35)[c]{$\flowiii \;\implies\; a_0 \preceq a_1 \preceq a_2 \preceq a_0$}
 		\end{picture}
		\iftwocol{\caption}{\figcaption}
 					{A geometry with a successor function $f: O\comp \to I\comp$\,, but
 					no causal flow.}
 		\label{fig:exampleNoFlow}
	\end{center}
\end{figure}

Given that the successor function of a causal flow for $(G,I,O)$ induces a path cover,
one might think of also trying to obtain a causal flow from the successor function of a
path cover. There is an obvious choice of binary relation for a successor function $f$\,:
\begin{definition}
	\label{dfn:naturalPreorder}
	Let $f$ be a successor function for $(G,I,O)$\,. The \emph{natural
	pre-order\footnote{A pre-order is a binary relation which is reflexive
	and transitive, but not necessarily antisymmetric.} $\preceq$
	for $f$} is the transitive closure on $V(G)$ of the conditions
	\begin{align}
		\label{eqn:naturalPreorder}
			x &\preceq x
			\;;\;\;
		&
			x &\preceq f(x)
			\;;\;\;
		&
			y \sim f(x) \;\;&\implies\;\; x \preceq y
			\onecol{\;\;,}
	\end{align}
	for all $x, y \in V(G)$\,.
\end{definition}
If $\preceq$ is a partial order, it will be the coarsest partial order such that
$(f, \preceq)$ is a causal flow. However, it is easy to construct geometries where
$\preceq$ is not a partial order. Figure~\ref{fig:exampleNoFlow} illustrates one
example. For any choice of successor function $f$ on this geometry, \flowiii~forces
either $a_0 \preceq a_1 \preceq a_2 \preceq a_0$ or $a_0 \succeq a_1 \succeq a_2 \succeq a_0$
to hold. Because $a_0$\,, $a_1$\,, and $a_2$ are distinct, such a relation $\preceq$
is not antisymmetric, so it isn't a partial order.

In the example above, we have a cycle of relationships induced by condition~\flowiii.
The following definitions characterize when such cycles of relationships occur.
\begin{definition}
	\label{dfn:viciousCircuit}
	Let $(G,I,O)$ be a geometry, and $\mF$ a family of directed paths in $G$\,. A walk
	$W = u_0 u_1 \cdots u_\ell$ is an \emph{influencing walk}\footnote{These are closely
	related to \emph{walks which alternate with respect to $\mF$}: see Section~\ref{sec:alternWalks}.}
	for $\mF$ if it is a concatenation of zero or more paths (called \emph{segments}
	of the influencing walk) of the following two types:
	\begin{itemize}
	\item
		$x y$\,, where $x \arc y$ is an arc of $\mF$\,;
	\item
		$x z y$\,, where $x \arc z$ is an arc of $\mF$ and $y z \in E(G)$\,.
	\end{itemize}
	A \emph{vicious circuit} for $\mF$ is a closed influencing walk for $\mF$ with at least
	one segment.
\end{definition}

\begin{citetheorem}[Lemma~9]{B06b}{Theorem}
	\label{thm:influencingNatlPreorder}
	Let $\mC$ be a path cover for $(G,I,O)$ with successor function $f$\,, and let $\preceq$
	be the natural pre-order of $f$\,. Then $x \preceq y$ if and only if there is an
	influencing walk for $\mC$ from $x$ to $y$\,.
\end{citetheorem}

Given that we want to forbid cycles of relationships for the natural pre-order $\preceq$\,,
we are then interested in the following restriction of path covers:

\begin{definition}
	\label{dfn:causalPathCover}
	A path cover $\mC$ for $(G,I,O)$ is a \emph{causal path
	cover} if $\mC$ does not have any vicious circuits in $G$\,.
\end{definition}

\begin{citetheorem}[Theorem~10]{B06b}{Theorem}
	\label{thm:graphth-charn}
	Let $(G,I,O)$ be a geometry with path cover $\mC$\,, $f$ be the successor function of
	$\mC$\,, and $\preceq$ be the natural pre-order for $f$\,. Then $\mC$ is a causal
	path cover if and only if $\preceq$ is a partial order, which occurs if and only if
	$(f,\preceq)$ is a causal flow for $(G,I,O)$\,.
\end{citetheorem}

By characterizing causal flows in terms of causal path covers, we can make use of the
following result:

\begin{citetheorem}[Theorem~11]{B06b}{Theorem}
	\label{thm:uniqueFamilyPaths}
	Let $(G,I,O)$ be a geometry such that $\abs{I} = \abs{O}$\,, and let $\mC$ be a path
	cover for $(G,I,O)$\,. If $\mC$ is a causal path cover, then $\mC$ is the only
	maximum collection of vertex-disjoint $I$ -- $O$ dipaths.
\end{citetheorem}

Then, if $\abs{I} = \abs{O}$ and $(G,I,O)$ has a causal flow, there is a unique maximum-size
collection of vertex-disjoint $I$ -- $O$ paths, and that collection is a causal path cover
which allows one to reconstruct a causal flow. Taking the contrapositive, if we can find
a maximum-size collection of vertex-disjoint paths from $I$ to $O$ which is not a causal
path cover, then $(G,I,O)$ does not have a causal flow.

\section{An efficient algorithm for finding a causal flow when $\abs{I} = \abs{O}$}
\label{sec:polyTimeAlg}

Using Theorems~\ref{thm:graphth-charn} and~\ref{thm:uniqueFamilyPaths} when
$\abs{I} = \abs{O}$\,, we can reduce the problem of finding a causal flow to finding
a maximum-size family of vertex-disjoint $I$--\;\!$O$ paths in $G$\,. Given such a family of paths
$\mF$\,, we may then verify that the resulting family forms a path cover for $G$\,, obtain
the successor function $f$ of $\mF$\,, and attempt to build a causal order compatible with
$f$\,. We illustrate how this may efficiently be done in this section.

\paragraph{Implementation details.}
\label{par:implDetails}
		
For the purpose of run-time analysis, I fix here conventions for the data structures
used to implement graphs, paths, and sets throughout the following algorithms.

\begin{itemize}
\item
	We will assume an implementation of graphs and digraphs using adjacency lists for
	each vertex $x$ (in the case of digraphs, using two separate lists for the arcs
	entering $x$ and those leaving $x$). Such an implementation can be easily performed
	in space $O(m)$\,, where $m$ is the number of arcs/edges, assuming a connected
	(di-)graph.\footnote{This holds, in particular, for graphs corresponding to one-way
	patterns implementing unitary operations which are not tensor-product decomposable.} 

\item
	Sets of vertices are considered to be implemented via arrays storing the characteristic
	function of the set. We may assume without loss of generality that these are also used to perform
	bounds-checking on arrays which are used to implement partial functions on $V(G)$\,, such as 
	successor functions $f: O\comp \to I\comp$\,.

\item
	Collections of vertex-disjoint di-paths $\mF$ in a graph $G$ will be implemented as
	a set $V(\mF)$ indicating for each $x \in V(G)$ whether $x$ is covered by $\mF$\,,
	and a digraph containing all of the arcs of $\mF$\,. As well, functions \prev\ and \next\
	will be defined for all vertices in $I\comp$ (respectively, $O\comp$) covered by $\mF$
	which returns the predecessor (respectively, successor) of a vertex covered by $\mF$\,.

	Throughout some of the algorithms below, a family of vertex-disjoint paths may be transformed
	into to a graph where a single vertex has out-degree $2$\,, but every other vertex has
	out-degree at most $1$\,, and every vertex has in-degree at most $1$\,. So long as these
	bounds are maintained, determining whether a vertex is covered by $\mF$\,, whether an
	arc is in $\mF$\,, and adding/deleting arcs from $\mF$ can be done in constant time.
	As well, the function $\prev$ will be well-defined so long as the in-degree of the graph
	representation of $\mF$ is bounded by $1$\,.

\end{itemize}

\subsection{Efficiently finding a path cover for $(G,I,O)$}
\label{sec:buildPathCover}

Given a geometry $(G,I,O)$\,, we are interested in obtaining a maximum-size family $\mF$ of
disjoint $I$--\;\!$O$ paths in $G$ in order to test whether it is a causal path cover.
This is known to be efficiently solvable.

Problems involving constructing collections of paths with some extremal property
in graphs are usually solved by reducing the problem to a problems of network flows
on digraphs: algorithms for such problems have been very well studied.
(Section~4.1 of~\cite{B06b} outlines an algorithm of this kind to find
a maximum-size family of disjoint $I$ -- $O$ paths.) However, in order to
present a solution which does not assume any background in graph-theoretic algorithms,
and also in order to reduce the number of auxiliary concepts involved in the solution,
I will present an algorithm not explicitly based on network flows.\footnote{The
solution presented here can be easily related to the solution via network flows,
but a small amount of additional work must be done in order to stay in the context of
collections of disjoint paths, rather than disjoint paths, cycles, and walks of length
$2$\,.} A dividend of such a presentation is that it highlights the relationship between
influencing walks and \emph{walks which alternate with respect to a collection of disjoint
paths}, which was alluded to in Definition~\ref{dfn:viciousCircuit}.

\subsubsection{Alternating and augmenting walks}
\label{sec:alternWalks}

\begin{definition}
	Let $I,O \subset V(G)$\,. A collection of vertex-disjoint paths from $I$ to $O$ 
	is \emph{proper} if its' paths intersect $I$ and $O$ only at their endpoints. 
\end{definition}

A collection of $k$ vertex-disjoint $I$ -- $O$ paths of is necessarily proper when
$\abs{I} = \abs{O} = k$\,. We would like to arrive at such a maximum-size collection
by producing successively larger proper collections of vertex-disjoint paths. To so so,
we will use results of graph theory pertaining to Menger's Theorem. The basic approach
present is outlined in Section~3.3 of~\cite{Diestel}.

\begin{definition}
	For a family $\mF$ of vertex-disjoint directed paths from $I$ to $O$\,, a walk
	$W = u_0 u_1 \cdots u_\ell$ in $G$ is said to be \emph{pre-alternating with respect
	to $\mF$} if the following hold for all $0 < j,k \le \ell$\,:
	\begin{romanum}
	\item
		\label{item:altern-i}
		$\mF$ does not contain $u_j \arc u_{j\!+\!1}$ as an arc;
	\item
		if $u_j = u_k$ and $j \ne k$\,, then $u_j$ is covered by $\mF$\,;
	\item
		\label{item:altern-iii}
		if $u_j$ is covered by $\mF$\,, then either $u_j \arc u_{j\!-\!1}$ or
		$u_{j\!+\!1} \arc u_j$ is an arc of $\mF$\,.
	\end{romanum}
	$W$ is said to be \emph{alternating with respect to $\mF$} if $W$ is pre-alternating
	with respect to $\mF$\,, and $u_0$ is an element of $I$ not covered by $\mF$\,. $W$ is an
	\emph{augmenting walk} for $\mF$ if $W$ alternates with respect to $\mF$\,, and $u_\ell \in O$\,.
\end{definition}

\begin{figure}[ht]
	\setlength\unitlength{0.16mm}
	\begin{center}
 		\begin{picture}(320,175)
 			\includegraphics[width=51mm]{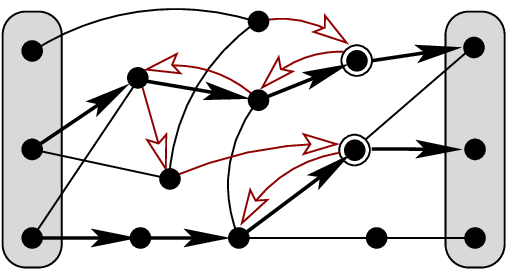}
			\puttext(-325,90)[r]{$I$}
			\puttext(5,90)[l]{$O$}
			\puttext(-150,170)[l]{$\scriptstyle u_0$}
			\puttext(-90,150)[l]{$\scriptstyle u_1$}
			\puttext(-145,100)[l]{$\scriptstyle u_2$}
			\puttext(-255,135)[l]{$\scriptstyle u_3$}
			\puttext(-240,50)[l]{$\scriptstyle u_4$}
			\puttext(-90,60)[l]{$\scriptstyle u_5$}
			\puttext(-165,10)[l]{$\scriptstyle u_6$}
 		\end{picture}
		\hspace{3cm}
 		\begin{picture}(320,175)
 			\includegraphics[width=51mm]{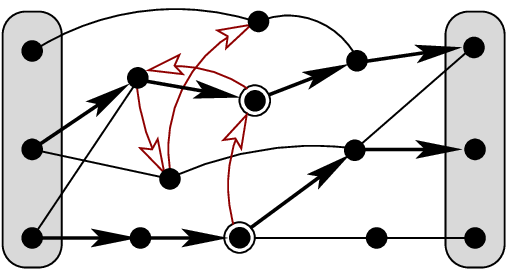}
			\puttext(-325,90)[r]{$I$}
			\puttext(5,90)[l]{$O$}
			\puttext(-160,5)[l]{$\scriptstyle u_0$}
			\puttext(-145,100)[l]{$\scriptstyle u_1$}
			\puttext(-255,135)[l]{$\scriptstyle u_2$}
			\puttext(-240,50)[l]{$\scriptstyle u_3$}
			\puttext(-150,170)[l]{$\scriptstyle u_4$}
 		\end{picture}
		\iftwocol{\caption}{\figcaption}
 					{Two examples of a walk $W$ (hollow arrows) which is pre-alternating with respect to a collection
					$\mF$ of vertex-disjoint paths from $I$ to $O$ (solid arrows). In both examples, circled
					vertices are \emph{entry points} of $W$ into $\mF$ (see Definition~\ref{dfn:properAlternWalk}).}
 		\label{fig:prealternWalk}
	\end{center}
\end{figure}

Figure~\ref{fig:prealternWalk} illustrates two pre-alternating walks for a family $\mF$ of vertex-disjoint
paths in a geometry $(G,I,O)$\,.
\vspace{-3ex}

The relationship between influencing walks and pre-alternating walks is most clear for a path cover $\mC$
of $(G,I,O)$\,, in which case an influencing walk for $\mC$ is the reverse of a walk which is pre-alternating
for $\mC$\,. As we will see in the next few pages, pre-alternating walks describe ways in which different families
of disjoint paths from $I$ to $O$ are related to each other: this is essentially the reason why a vicious
circuit (i.e. a closed influencing walk) exists for a path cover whenever there is a second family of
disjoint $I$ -- $O$ paths of the same size.

First, we will show that augmenting walks for $\mF$ are always present if $\abs{\mF} <
\abs{I} = \abs{O}$\,, and if there is a family of vertex-disjoint $I$ -- $O$ paths
of size $k$\,:

\begin{theorem}
	\label{thm:freeAugmentingWalk}
	Let $G$ be a graph, and $I, O \subset V(G)$ with $\abs{I} = \abs{O} = k$\,. Let $\mF$
	be a collection of vertex-disjoint $I$ -- $O$ paths with $\abs{\mF} < k$\,, and $i \in I$
	be a vertex not covered by $\mF$\,. If there is a collection $\mC$ of vertex-disjoint
	dipaths from $I$ to $O$ with $\abs{\mC} = k$\,, then there is an augmenting walk $W$ for $\mF$
	starting at $i$ which traverses each edge of $G$ at most once, and where $i$ is the
	only input vertex in $W$ not covered by $\mF$\,.
\end{theorem}
\begin{proof}
	Suppose $G$ contains a collection $\mC$ of $k$ vertex-disjoint $I$ -- $O$ dipaths, let
	$\mF$ be some proper collection of vertex-disjoint $I$ -- $O$ dipaths of size less than
	$k$\,, and let $I' \ne \vide$ be the set of input vertices not covered by $\mF$\,.
	Let us say that a vertex $v \in V(G)$ is an \emph{incidence point} of $\mC$ and $\mF$
	if $v$ is covered by both $\mC$ and $\mF$\,, and there is a vertex $w$ which is
	adjacent to $v$ in a path of $\mC$ but which is not adjacent to $v$ in any path of
	$\mF$\,.	Let $\sI$ be the set of incidence points of $\mC$ and $\mF$\,, and let $\sG$ be a
	di-graph with $V(\sG) = I' \union \sI \union O$\,, and $(x \arc y) \in A(\sG)$ for
	$x,y \in V(Q)$ if one of the following applies:
	\begin{itemize}
	\item
		there exists a vertex $z \in \sI$ such that
		\begin{romanum}
		\item
			$x$ and $z$ lie on a common path $P$ in $\mC$\,, where $z$ is the next incidence
			point in $P$ after $x$\,, and
		\item
			$y$ and $z$ lie on a common path $P'$ in $\mF$\,, where $z$ is the next incidence
			point in $P'$ after $y$\,;
		\end{romanum}
		
	\item
		$x$ and $y$ lie on a common path $P$ in $\mC$\,, there are no incidence points on $P$
		after $x$\,, and $y \in O$\,.
	\end{itemize}

	\begin{figure}[ht]
	~\hfill
	\begin{minipage}[t]{0.45\textwidth}
		\setlength\unitlength{0.16mm}
		\begin{center}
 			\begin{picture}(320,175)
 				\includegraphics[width=51mm]{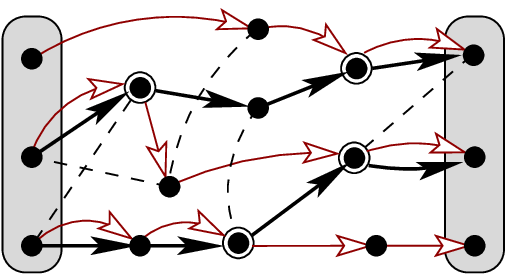}
				\puttext(-325,90)[r]{$I$}
				\puttext(5,90)[l]{$O$}
 			\end{picture}
			\iftwocol{\caption}{\figcaption}
 						{Two families of vertex-disjoint $I$ -- $O$ paths in a graph: one family $\mC$ with $k$ paths
						(hollow arrows), and one family $\mF$ with $< k$ paths (solid arrows). Circled vertices are the
						incidence points of $\mC$ and $\mF$\,. Dashed lines are the other edges of the graph.}
 			\label{fig:augWalkInfluencing}
		\end{center}
	\end{minipage}
	\hfill
	\begin{minipage}[t]{0.45\textwidth}
		\setlength\unitlength{0.16mm}
		\begin{center}
 			\begin{picture}(320,175)
 				\includegraphics[width=51mm]{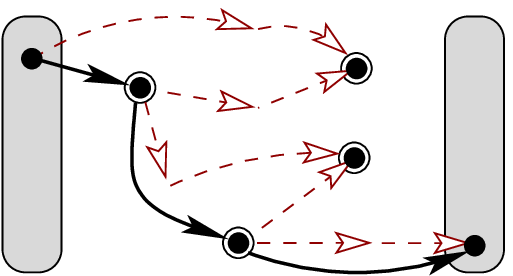}
				\puttext(-325,90)[r]{$I$}
				\puttext(5,90)[l]{$O$}
 				\puttext(-160,170)[l]{$\scriptstyle \mC$}
 				\puttext(-160,120)[l]{$\scriptstyle \mF$}
 				\puttext(-205,75)[l]{$\scriptstyle \mC$}
 				\puttext(-155,50)[l]{$\scriptstyle \mF$}
 				\puttext(-85,32)[l]{$\scriptstyle \mC$}
 			\end{picture}
			\iftwocol{\caption}{\figcaption}
 						{The digraph $\sG$ obtained by applying the construction above to Figure~\ref{fig:augWalkInfluencing}. 
						Dashed arrows represent the edges from path segments belonging to either $\mC$ or $\mF$ in the
						original graph; thick black arrows are the actual arcs of $\sG$\,, which are induced
						by those path segments.}
 			\label{fig:augWalkInfluencingSkel}
		\end{center}
	\end{minipage}
	\hfill~
	\end{figure}

	Because both $\mC$ and $\mF$ are vertex-disjoint collections of paths, it is easy to show
	that the maximum in-degree and out-degree of $\sG$ are both $1$\,. Thus, $\sG$ consists of
	vertex-disjoint di-cycles, walks of length $2$\,, isolated vertices, and directed paths.

	Because each $v \in I'$ is not covered by a path of $\mF$\,, and is not preceded by any
	vertices in it's respective path of $\mC$\,, it has in-degree $0$ in $\sG$\,. Then, each
	element of $I'$ is at the beginning of a maximal dipath in $\sG$\,. Furthermore, each vertex
	in $v \in I' \union \sI$ has out-degree $1$\,: if $P \in \mC$ is the path covering $v$\,,
	either there are no incidence vertices after $v$ on $P$\,, in which case there is an arc
	$v \arc y$ for the vertex $y \in O$ at the end of $P$\,; or if we let $z \in \sI$ be the
	first incidence vertex following $v$ on $P$\,, we will have $z \notin I$\,, in which case
	there will be an incidence vertex $w$ which precedes $z$ on some path of $\mF$\,,
	because all input vertices covered by $\mF$ are incidence points. Thus, any maximal di-path
	in $\sG$ must end in $O$\,. Then, for each $i \in I'$\,, there is a dipath from $i$ to
	some element of $O$ in the graph $\sG$\,.

	Consider any vertex $i \in I'$\,, and let $u_0 \arc u_1 \arc \cdots \arc u_\ell$
	be the dipath in $\sG$ from $i$ to $O$\,. Let $P \in \mC$ and $P' \in \mF$ be the paths
	containing $u_{\ell - 1}$\,: from $(u_{\ell - 2} \arc u_{\ell - 1}) \in A(\sG)$\,, we know
	that there is an incidence vertex after $u_{\ell - 1}$ in the path $P'$\,. Note that $u_\ell$
	is either not covered by any path of $\mF$\,, or it occurs at the end of a path of $\mF$
	and is not followed by any vertices on that path; then, $(u_{\ell - 1} \arc u_\ell) \in
	A(\sG)$ implies that there are no incidence points on $P$ after $u_{\ell - 1}$\,. Then,
	the arcs leaving $u_{\ell - 1}$ in $P$ and $P'$ must be different: the fact that no incidence
	point follows $u_{\ell - 1}$ in $P$ then implies that no path of $\mF$ intersects $P$
	after $u_{\ell - 1}$\,. In particular, $u_\ell$ is not covered by $\mF$\,.

	Because $u_0 \in I'$ and $u_\ell \in O$ are both not covered by $\mF$\,, we may construct
	an augmenting walk $W$ for $\mF$ in the original graph $G$\,, as follows. If $\ell = 0$\,,
	we let $W$ be the trivial path on $u_0$\,, which is an augmenting walk for $\mF$\,.
	Otherwise:
	\begin{itemize}
	\item
		For each $j \in [\ell - 1]$\,, let $v_j$ be the next incidence point after $u_j$ on
		the path $P_j \in \mC$ containing $u_j$\,. (This $v_j$ will then also be the next
		incidence point after $u_{j\!+\!1}$ on the path $P'_j \in \mF$ containing
		$u_{j\!+\!1}$\,.)
	\item
		Let $\tilde{P}_j$ be the segment of $P_j$ from $u_j$ to $v_j$\,, and $\tilde{P}'_j$
		be	the \emph{reverse} of the segment of $P'_j$ from $u_{j\!+\!1}$ to $v_j$\,.
	\item
		Finally, let $\tilde{P}_{\ell\!-\!1}$ be the path segment in $\mC$ from
		$u_{\ell\!-\!1}$ to $u_\ell$\,.
	\end{itemize}
	Then, define $W \;=\; u_0 \tilde{P}_0 v_0 \tilde{P}'_0 u_1 \tilde{P}_1 \cdots u_{\ell\!-\!1}
	\tilde{P}_{\ell\!-\!1} u_\ell$\;. We may show that $W$ is an augmenting walk for $\mF$\,:
	\begin{romanum}
	\item
		Each path $\tilde{P}_j$ is internally disjoint from $\mF$\,, because they are
		sub-paths of elements of $\mC$\,, and do not contain any incidence points in their
		interiors. Then, none of the arcs of $\tilde{P}_j$ are arcs of $\mF$ for any $j \in
		[\ell]$\,. Also, all of the arcs of the paths $\tilde{P}'_j$ are the reverse of arcs of
		$\mF$\,: they do not contain arcs of $\mF$ either. Then, none of the arcs of $W$
		are arcs of $\mF$\,.
	\item
		Because $u_0 \arc \cdots \arc u_\ell$ is a directed path in $\sG$\,, we have $u_j \ne
		u_k$\,. Because each path $\tilde{P}_j$ and $\tilde{P}'_j$ is uniquely determined by
		$u_j$ for $j \in [\ell - 1]$\,, those sequences of vertices can also occur only once
		each. Each interior vertex of $\tilde{P}_j$ or $\tilde{P}'_j$ can only
		occur in a single path of $\mC$ or $\mF$\,, between two consecutive elements of
		$I' \union \sI \union O$ on that path\,: then, because each segment $\tilde{P}_j$
		and $\tilde{P}'_j$ only occur once in $W$\,, each interior vertex of those segments
		also occurs only once in $W$\,.

		Thus, if any vertex $x$ occurs more than once in $W$\,, $x$ must be an end point of
		some path $\tilde{P}_j$ or $\tilde{P}'_j$\,. Aside from $\tilde{P}_0$ and
		$\tilde{P}_{\ell - 1}$\,, both end-points of each such segment has in-degree $1$
		and out-degree $1$\,, so they cannot be elements of either $I'$ or $O$\,. Then,
		any vertex which occurs more than once in $W$ is an element of $\sI$\,, and is
		therefore covered by $\mF$\,.
	\item
		The only points in $W$ which are covered by $\mF$ are the vertices of the paths
		$\tilde{P}_j$ for $j \in [\ell - 1]$\,, which are all at the beginning or the end
		of arcs in $W$ which are the reverse of arcs of $\mF$\,.
	\end{romanum}
	Thus, $W$ is an augmenting walk for $\mF$\,. Furthermore, because each edge of $G$ is
	contained in at most one segment $\tilde{P}_j$ or $\tilde{P}'_j$\,, each edge occurs at
	most once in $W$\,. Finally, because elements of $I'$ have in-degree $0$ in $\sG$ and
	do not occur in the segments $\tilde{P}_j$ or $\tilde{P}'_j$\,, any input vertices other
	than $i = u_0$ which occur on $W$ must be incidence points, which means they are covered by
	$\mF$\,. Thus, there is a proper augmenting path for $\mF$ of the desired type starting
	at $i \in I$\,.
\end{proof}

\vspace{-1ex}
The above Theorem illustrates how we can build an augmenting walk for $\mF$ from a collection
of disjoint $I$ -- $O$ paths which covers $I$ and $O$\,. If we impose restrictions on the type
of augmenting walk we consider, we may also efficiently do the reverse. The restriction we are
interested in is the following:
\vspace{-1ex}

\begin{definition}
	\label{dfn:properAlternWalk}
	Let $W = u_0 u_1 \cdots u_\ell$ be a walk which which is pre-alternating with respect to
	$\mF$\,.
	\vspace{-2ex}
	\begin{itemize}
	\item
		An \emph{entry point} of $W$ into $\mF$ is a vertex $u_j$ which is covered by $\mF$\,,
		where either $j = 0$ or $u_j \arc u_{j-1}$ is not an arc of $\mF$\,.

	\item
		The walk $W$ is \emph{monotonic} if, for every path $P \in \mF$ and for
		any indices $0 \le h < j < \ell$ such that $u_h$ and $u_j$ are both entry points for $W$
		into $\mF$ which lie on $P$\,, $u_h$ is closer to the initial point of $P$
		than than $u_j$ is.
	\item
		$W$ is a \emph{proper} pre-alternating walk if $W$ traverses each edge at most once,
		each input vertex in $W$ (except possibly $u_0$) is covered by $\mF$\,, and $W$ is monotonic.
	\end{itemize}
\end{definition}

We will be most interested in proper augmenting walks, which are useful in increasing the size
of proper collections of $I$ -- $O$ paths. The sort of augmenting walk that is guaranteed
by Theorem~\ref{thm:freeAugmentingWalk} is almost a proper augmenting walk, and merely
lacks a guarantee of monotonicity. However, the following Lemma shows that we lose no generality
in imposing monotonicity as a condition:

\begin{lemma}
	\label{lemma:wlogMonotonic}
	Let $G$ be a graph, and $I, O \subset V(G)$\,. Let $\mF$	be a collection of vertex-disjoint
	$I$ -- $O$ paths, and let $W$ be an augmenting walk for $\mF$ from $i \in I$ to $\omega \in O$\,.
	Then there is a monotonic augmenting walk $\sW$ for $\mF$ from $i$ to $\omega$\,.
\end{lemma}
\begin{proof}
	Let $W$ be given by $W = u_0 \cdots u_\ell$\,, where $u_0 = i$ and $u_\ell = \omega$\,. For any
	path $P \in \mF$\,, and two entry points $u_h$ and $u_j$ of $W$ into $\mF$\,, let us say that
	$(u_h, u_j)$ is a \emph{reversed pair} if $h < j$ but $u_j$ is closer to the initial point
	of $P$ than $u_h$\,. We will produce a monotonic augmenting walk by recursively reducing
	the number of reversed pairs of $W$\,.
	\begin{itemize}
	\item
		If $W$ has no reversed pairs, then $W$ is already monotonic, in which case we may let
		$\sW = W$\,.
	\item
		Suppose that $(u_h, u_j)$ is a reversed pair of $W$\,. Then $h < j$\,, but $u_j$ is
		closer than $u_h$ to the initial point of the path $Q \in \mF$ which covers both of
		them. Note that $u_\ell = \omega$ is not covered by $\mF$\,, and so is not on the
		path $Q$\,: then, let $j' \in [\ell]$ be the smallest index such that $u_{j'+1}$
		is not on $Q$\,. Let $Q = q_0 \cdots q_a q_{a+1} \cdots q_{b-1} q_b \cdots q_m$\,,
		where $q_a = u_{j'}$ and $q_b = u_h$\,. Then, let
		\begin{align*}
			W' = u_0 \cdots u_{h-1} q_b q_{b-1} q_{b-2} \cdots q_{a+1} q_a u_{j'+1} \cdots u_\ell	\;.
		\end{align*}
		From the fact that $W$ is an augmenting walk for $\mF$, it is easy to show that $W'$ is
		also an augmenting walk for $\mF$\,. As well, the entry 	points of $W'$ into $\mF$ are a
		subset of the entry points of $W$ into $\mF$\,, in which case the reversed	pairs of $W'$ are
		also a subset of the reversed pairs of $W$\,; and $W'$ does not have $(u_h, u_j)$ as a reversed
		pair. Then, $W'$ has strictly fewer reversed pairs than $W$\,.
	\end{itemize}
	Because $W$ is a finite walk, it can have only finitely many reversed pairs; then, by recursion,
	we may construct a monotonic augmenting walk $\sW$ for $\mF$ from $i$ to $\omega$\,.
\end{proof}

\begin{corollary}
	\label{cor:freeProperAugmentingWalk}
	Suppose $\abs{I} = \abs{O} = k$\,, $\mF$ a proper collection of vertex-disjoint $I$ -- $O$ paths in
	$G$ with $\abs{\mF} < k$\,, and let $i \in I$ be a vertex not covered by $\mF$\,. If there is a
	collection $\mC$ of vertex-disjoint dipaths from $I$ to $O$ with $\abs{\mC} = k$\,, then there
	is a proper augmenting walk $W$ for $\mF$ starting at $i$\,.
\end{corollary}
\begin{proof}
	Theorem~\ref{thm:freeAugmentingWalk} and Lemma~\ref{lemma:wlogMonotonic}.
\end{proof}

For proper augmenting walks, the reason for requiring that no edge is traversed twice is essentially
to help construct efficient algorithms for finding them, which we consider later. The requirements 
that the only input vertex in the walk which is not covered by $\mF$\,, and that it be monotonic,
are essentially chosen to allow us to use augmenting walks to increase the size of a \emph{proper}
collection of vertex-disjoint paths to cover exactly one more input vertex. We may do this using
the following operation:

\begin{definition}
	\label{dfn:augmentCollectionPaths}
	Let $\mF$ be a proper collection of vertex-disjoint $I$ -- $O$ dipaths in $G$\,, and $W$ be
	a proper augmenting walk for $\mF$\,. Then, $\mF \oplus W$ denotes the collection of directed
	paths which are formed by those arcs $x \arc y$ which belong either to $W$ or a path of
	$\mF$\,, and for which $y \arc x$ is not an arc of either $W$ or $\mF$\,.
\end{definition}
\begin{figure}[ht]
	\setlength\unitlength{0.16mm}
	\begin{center}
		\begin{minipage}[c]{55mm}
 		\begin{picture}(320,175)
 			\includegraphics[width=51mm]{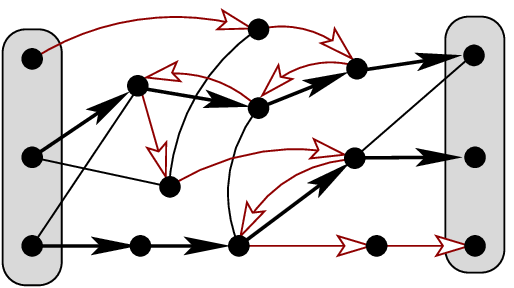}
			\puttext(-325,90)[r]{$I$}
			\puttext(5,90)[l]{$O$}
 		\end{picture}
		\end{minipage}
		$\qquad\mapsto\qquad\quad$
		\begin{minipage}[c]{55mm}
 		\begin{picture}(320,175)
 			\includegraphics[width=51mm]{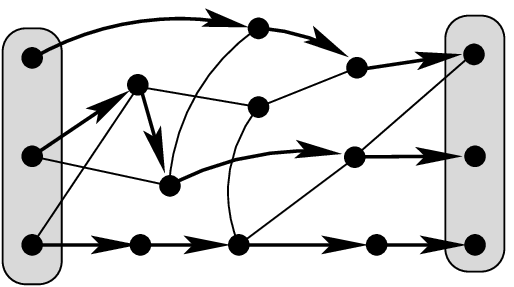}
			\puttext(-325,90)[r]{$I$}
			\puttext(5,90)[l]{$O$}
 		\end{picture}
		\end{minipage}
		\iftwocol{\caption}{\figcaption}
 					{On the left: an example of a proper collection $\mF$ of vertex disjoint $I$ -- $O$ paths
					(solid arrows) with a proper augmenting walk $W$ for $\mF$ (hollow arrows). On the right: the
					augmented collection of paths $\mF \oplus W$\,.}
 		\label{fig:augmentCollectionPaths}
	\end{center}
\end{figure}

The collection $\mF \oplus W$ described above is formed by the usual procedure for augmenting
a network-flow with an augmenting path: one can think of forming $\mF \oplus W$ by ``adding''
together the arcs of $\mF$ and $W$\,, and ``cancelling'' them whenever they point in opposite directions
on a single edge.

\begin{lemma}
	\label{lemma:augmentCollectionPaths}
	Let $\mF$ be a proper collection of vertex-disjoint $I$ -- $O$ dipaths in $G$\,, and $W$
	be a proper augmenting walk for $\mF$\,. Then $\mF \oplus W$ is a proper collection of
	vertex-disjoint $I$ -- $O$ dipaths, with $\abs{\mF \oplus W} = \abs{\mF} + 1$\,; and the
	input vertices covered by $\mF \oplus W$ are those covered by $\mF$ and $W$ together.
\end{lemma}

\begin{proof}
	We induct on the number of times $r$ that the walk $W$ intersects the paths of $\mF$\,.
	If $r = 0$\,, then $\mF \oplus W = \mF \union \brace{W}$\,, and the inputs covered by
	$\mF \oplus W$ are clearly those covered by $\mF$ or by $W$\,. Otherwise, suppose that
	the proposition holds for all cases where the augmenting walk intersects the paths
	of its' respective collection fewer than $r$ times.
	
	Let $W$ be given by $W = u_0 u_1 \cdots u_a u_{a\!+\!1} \cdots u_{b\!-\!1} u_b \cdots
	u_\ell$\,, where none of the points $u_j$ are covered by $\mF$ for $j \in [a]$\,, and
	where $u_{j\!+\!1} \arc u_j$ is an arc of $\mF$ for all $a \le j < b$\,.
	Let $Q \in \mF$ be the path containing $u_a$ through $u_b$\,: in particular,
	let $Q = q_0 q_1 \cdots q_c q_{c\!+\!1} \cdots q_{d\!-\!1} q_d \cdots q_m$\,,
	where $q_c = u_b$ and $q_d = u_a$\,. Then, we may define
	\begin{align*}
			Q'
		\;\;=&\;\;
			u_0 u_1 \cdots u_a q_{d\!+\!1} \cdots q_m	\;\;,
		&
			W'
		\;\;=&\;\;
			q_0 q_1 \cdots q_c u_{b\!+\!1} \cdots u_\ell \;\;:
	\end{align*}
	then $Q' \in \mF \oplus W$\,, and $W'$ is an augmenting walk for $\mF' = \paren{\mF \setminus Q}
	\union \brace{Q'}$ which intersects the paths of $\mF'$ fewer than $r$ times. Because $\mF$
	is a proper collection of vertex-disjoint $I$ -- $O$ paths, $Q'$ only intersects
	$I$ and $O$ at its' endpoints, and $Q'$ does not intersect any paths of $\mF \setminus Q$\,,
	$\mF'$ is proper. Similarly, because $Q$ only intersects $I$ at $q_0$ and because $W$ only
	intersects $I$ at $u_0$ and at input vertices covered by $\mF$\,, $W'$ does not cover any
	inputs except those covered by $\mF$\,. Because $W$ doesn't traverse any edges twice,
	and all of the other entry points $q_h$ of $W$ into $\mF$ on the path $Q$ have $h > c$ by the
	monotonicity of $W$\,, $W'$ itself does not traverse any edge twice. Finally, all of the
	entry points of $W$ into $\mF$ are also entry points of $W'$ into $\mF'$\,, except for
	$u_a$\,: all the other are left unaffected, including the order in which they occur.
	Then $W'$ is monotonic, so that $W'$ is a proper augmenting walk for $\mF'$\,.

	By the induction hypothesis, $\mF' \oplus W'$ is a proper collection of vertex-disjoint paths
	from $I$ to $O$\,, with $\abs{\mF' \oplus W'} = \abs{\mF'} + 1 = \abs{\mF} + 1$\,. Also by 
	induction, the input vertices covered by $\mF' \oplus W'$ are those covered by $\mF'$ or by
	$W'$\,. Because $W'$ covers the input $q_0$\,, and $\mF'$ covers all inputs covered by $W$
	or by $\mF$ except for $q_0$\,, $\mF' \oplus W'$ then covers all vertices covered by $\mF$
	or by $W$\,. Finally, note that the set of arcs from $\mF'$ and $W'$ together only differs
	from the set of arcs from $\mF$ and $W$ together by the absence of the arcs $u_j \arc u_{j+1}$
	from $W$ and the arcs $u_{j+1} \arc u_j$\,, for $a \le j < b$\,, which oppose each other.
	We then have $\mF' \oplus W' = \mF \oplus W$\,: thus, $\abs{\mF \oplus W} = \abs{\mF} + 1$\,,
	and $\mF \oplus W$ covers the input vertices covered either by $\mF$ or by $W$\,.
\end{proof}

\subsubsection{An efficient algorithm for finding a proper augmenting walk}

Algorithm~\ref{alg:augmentSearch} determines if a vertex supports a suitable
proper pre-alternating walk $W$ with respect to $\mF$\,, and compute $\mF \oplus W$ if
one is found. Using it, we may find proper augmenting walks for $\mF$ by performing
a depth-first search along proper alternating walks $W$ for $\mF$ in an attempt to find one
which ends in $O$\,. 

\begin{algorithm}[ht]
	\mycaption
		{\AugmentSearch(G, I, O, \mF, \iter, \visited, v)}
		{searches for an output vertex along pre-alternating walks for $\mF$ starting at $v$\,, subject
		to limitations on the end-points of the search paths.}
	\label{alg:augmentSearch}
	\begin{algorithmic}[1]
	\REQUIRE $(G,I,O)$ is a geometry.
	\REQUIRE $\mF$ is a specification for a vertex-disjoint family of $I$--$O$ paths.
	\REQUIRE $\iter$ is a positive integer.
	\REQUIRE $\visited$ is an array $V(G) \to \N$\,.
	\REQUIRE $v \in V(G)$\,.
	\medskip
	
	\setlength\commentskip{48ex}
	\STATE $\visited(v) \gets \iter$\,;
			\label{line:augment-markVtx}
	
	\smallskip

	\IFTHEN {$v \in O$}
		\label{line:augmentSearch-condSucceed}
		\algorithmicreturn $(\mF,\visited,\success)$.

	\smallskip

	\IF {$v \in V(\mF)$ \AND\ $v \notin I$ \AND\ $\visited(\prev(\mF,v)) < \iter$}
		\label{line:augmentSearch-condReverse}
		\STATE $(\mF, \visited, \status) \gets \AugmentSearch(G,I,O,\mF,\iter,\visited,\prev(\mF,v))$\,;
			\label{line:augmentSearch-visitReverse}
		\IF 
			 {$\status = \success$}
			\label{line:augmentSearch-preSuccessReverse}
			\STATE $\mF \gets \RemoveArc(\mF,\, \prev(\mF,v) \arc v)$\,;
			\RETURN $(\mF, \visited, \success)$\,.
					\label{line:augmentSearch-successReverse}
		\ENDIF
	\ENDIF

	\smallskip

	\setlength\commentskip{52ex}
	\FORALL {$w \sim v$}
		\label{line:augmentSearch-forLoop}
		\IF {$\visited(w) < \iter$ \AND\ $w \notin I$ \AND\ $(v \arc w) \notin A(\mF)$}
		\label{line:augmentSearch-condIter}
			\IF {$w \notin V(\mF)$}
				\label{line:augmentSearch-condOther}
				\setlength\commentskip{35ex}
				\STATE $(\mF, \visited, \status) \gets \AugmentSearch(G,I,O,\mF,\iter,\visited,w)$\,;
					\label{line:augmentSearch-visitOther}
				\IF 
					 {$\status = \success$}
					\STATE $\mF \gets \AddArc(\mF,\, v \arc w)$\,;
					\RETURN $(\mF, \visited, \success)$\,.
						\label{line:augmentSearch-successOther}
				\ENDIF
			\ELSIF 
					  {$\visited(\prev(\mF,w)) < \iter$}
					\label{line:augmentSearch-condEntryPoint}
				\STATE $(\mF, \visited, \status) \gets \AugmentSearch(G,I,O,\mF,\iter,\visited,\prev(\mF,w))$\,;
					\label{line:augmentSearch-visitEntryPoint}
				\setlength\commentskip{41ex}
				\IF 
					 {$\status = \success$}
					\STATE $\mF \gets \RemoveArc(\mF,\, \prev(\mF,w) \arc w)$\,;
					\STATE $\mF \gets \AddArc(\mF,\, v \arc w)$\,;
					\RETURN $(\mF, \visited, \success)$\,.
						\label{line:augmentSearch-successEntryPoint}
				\ENDIF
			\ENDIF
		\ENDIF
		\setlength\commentskip{35ex}
	\ENDFOR

	\smallskip

	\setlength\commentskip{35ex}
	\RETURN $(\mF, \visited, \fail)$\,.
		\label{line:augmentSearch-fail}
	\end{algorithmic}
\end{algorithm}

\begin{theorem}
	\label{thm:augmentSearch}
	Let $(G,I,O)$ be a geometry with $\abs{I} = \abs{O} = k$\,, $\mF$ a proper collection of 
	fewer than $k$ vertex-disjoint paths from $I$ to $O$\,, $\iter$ a positive integer, $i \in I$ a
	vertex not covered by $\mF$\,, and $\visited: V(G) \to \N$ with $\visited(x) < \iter$ for
	all $x \in V(G)$\,. Then \AugmentSearch\ halts on input $(G,I,O,\mF,\iter,\visited,i)$\,.
	Furthermore, let $(\bar\mF,\, \bar\visited,\, \status) = \AugmentSearch(G,I,O,\mF,\iter,\visited,i)$\,.
	\begin{romanum}
	\item
		\label{item:augmentSearch-i}
		If $\status = \fail$\,, then there are no proper augmenting walks for $\mF$ starting at $i$\,;
	\item
		\label{item:augmentSearch-ii}
		If $\status = \success$\,, then $\bar\mF$ is a proper family of vertex-disjoint $I$ -- $O$
		paths of size $\abs{\mF} + 1$ which covers $i$ and all input vertices covered by $\mF$\,,
		and $\bar\visited(x) \le \iter$ for all $x \in V(G)$\,.
	\end{romanum}
\end{theorem}
\begin{proof}
	Let $G$, $I$, $O$, $\mF$, and $\iter$ be fixed as above. Throughout the proof, we will
	consider chains of recursive calls to \AugmentSearch\,. One invocation of \AugmentSearch\
	is the \emph{daughter} of a second invocation if the first invocation was performed as a step
	of the second invocation; if one invocation is related to a second invocation by a sequence
	of daughter-relationships, we will call the second
	invocation a \emph{descendant} of the first. 

	At any stage in a particular invocation of \AugmentSearch\,, we will refer to the ordered pair
	$(\visited, v)$ as the \emph{data pair} of the invocation, where $v$ is the final parameter of
	the input, and $\visited$ the second last parameter, including any changes which have been made
	to it during the invocation. (Though the input parameters of \AugmentSearch\ include $G$,
	$I$, $O$, $\mF$, and $\iter$, we will occasionally refer to data pairs as the input of
	an invocation of \AugmentSearch\,.) When an invocation of \AugmentSearch\ has a data pair
	$(\visited,v)$ and makes a daughter invocation, we may describe that invocation as being
	``daughter invocation for $(\visited, v)$''\,; similarly, a daughter invocation for
	$(\visited, v)$ or the descendant of one is a ``descendant invocation for $(\visited, v)$''.

	We define a \emph{probe walk} $\sW$ for an ordered pair $(\visited, v)$ to be a proper
	pre-alternating walk starting at $v$ such that, for all vertices $x$ in the walk, $\visited(x)
	= \iter$ only if $x$ is at the beginning of $\sW$ and $x \in O$ only if $x$ is covered by
	$\mF$ or $x$ is at the end of $\sW$\,. Then, we let $R(\visited, v)$ be the set of vertices $x \in V(G)$
	which end-points of probe walks for $(\visited, v)$\,. We will reduce the problem
	of determining whether there is a proper augmenting path for $\mF$ passing through $v$
	to a question of the existence of whether there is an output vertex in $R(\visited, v)$\,, 
	for $\visited$ restricted in a manner described below.

	A \emph{canonical walk} for a data pair $(\visited, v)$ is a proper alternating walk $W$ with
	respect to $\mF$\,, such that the following all hold:
	\begin{romanum}
	\item
		$\visited(x) \le \iter$ for all $x \in V(G)$\,.

	\item
		$v$ is the end-point of $W$\,.

	\item
		for all vertices $x$ on $W$\,, if $\visited(x) < \iter$\,, then either $x = v$\,,
		or $x$ occurs exactly once in $W$ and is an entry point of $W$ into $\mF$\,.
	
	\item
		for any path $P$ of $\mF$\,, and $x \in V(P)$ which is not in $W$\,,
		$\visited(x) < \iter$ if and only if either
		\begin{inlinum}
		\item~there is no entry point of $W$ after $x$ on the path $P$\,, or
		\item~there is exactly one entry point $p$ of $W$ after $x$ on the path $P$\,,
				and $v$ lies on $P$ strictly between $x$ and $p$\,.
		\end{inlinum}
	\end{romanum}
	A data pair $(\visited, v)$ is itself \emph{canonical} if it has a canonical walk.
	We will be interested in the behaviour of \AugmentSearch\ on canonical inputs. (Note that the
	input described in the statement of the Theorem is a special case.) We will show that \AugmentSearch\
	essentially performs a depth-first traversal of $R(\visited, v)$ along probe walks for
	$(\visited, v)$ in an attempt to find an output vertex. If it succeeds, it has traversed
	a proper augmenting walk $\bar\sW$ for $\mF$\,, and can construct $\mF \oplus \bar\sW$\,.

	Suppose $W$ is a canonical walk for a data pair $(\visited, v)$\,. It is easy to show that if we extend
	$W$ to a longer walk $W' = W v w_1 w_2 \cdots w_N$ for some $N \ge 1$, and $W'$ is a canonical walk
	for some data pair $(\visited^\ast, w_N)$\,, then $W$ is a canonical walk for $(\visited^\ast, v)$\,.
	We will use this fact frequently in the two Lemmas below.

	\let\oldQed\qed
	\renewcommand\qed{\hfill$\Diamond$}
	\newcounter{tmpTheorem}
	\setcounter{tmpTheorem}{\value{theorem}}
	\setcounter{theorem}{0}
	\renewcommand\thetheorem{\ref{thm:augmentSearch}-\arabic{theorem}}

	\begin{lemma}
		\label{sublemma:augmentFail}
		Suppose that $(\visited, v)$ has a canonical walk $W$\,. If $R(\visited, v)$ does not contain
		any output vertices, \AugmentSearch\ halts on input data $(\visited, v)$\,, with output value
		$(\mF,\, \bar\visited,\, \fail)$\,; where $\bar\visited$ differs from $\visited$ only in that
		$\bar\visited(x) = \iter$ for all $x \in R(\visited, v)$\,, and where $(\bar\visited, v)$ also
		has the canonical walk $W$\,.
	\end{lemma}
	\begin{proof}
		We will proceed by induction on the length $\ell$ of the longest probe walk for $(\visited, v)$\,.
		Regardless of the value of $\ell$\,, line~\ref{line:augment-markVtx} transforms the data pair
		$(\visited, v)$ to $(\visited^{(1)}, v)$\,, where $\visited^{(1)}$ differs from $\visited$ in
		that $\visited^{(1)}(v) = \iter$\,; then, any canonical walk for $(\visited, v)$ is also a canonical
		walk for $(\visited^{(1)}, v)$\,. As well, it cannot be that $v \in O$\,: then the condition 
		on line~\ref{line:augmentSearch-condSucceed} will not be satisfied.         

		If $\ell = 0$\,, the condition on lines~\ref{line:augmentSearch-condReverse} cannot be satisfied,
		and the condition of line~\ref{line:augmentSearch-condIter} is not satisfied by any neighbor
		$w \sim v$. Then, line~\ref{line:augmentSearch-fail} will ultimately be executed, returning
		$(\mF, \visited^{(1)}, \fail)$\,. Because $R(\visited, v) = \brace{v}$\,, the proposition
		holds in this case.

		Otherwise, suppose $\ell > 0$\,, and that the proposition holds for canonical data pairs whose probe
		walks all have length less than $\ell$\,. Consider the vertices which may be the subject of a daughter
		invocation of \AugmentSearch:
		\renewcommand\labelenumi{\emph{\arabic{enumi}}.}
		\begin{enumerate}
		\item
			If $v \notin I$ and $v$ is covered by $\mF$\,, and $z$ is the predecessor of $v$ in the paths
			of $\mF$\,, then $(\visited, v)$ has probe walks starting with the arc $v \arc z$ if and only
			if $\visited^{(1)}(z) = \visited(z) < \iter$\,. If this holds, then a daughter invocation of
			\AugmentSearch\ with input data $(\visited^{(1)}, z)$ is performed.

			In this case, note that $(\visited^{(1)}, z)$ has probe walks ending in $O$ only if $(\visited, v)$
			does; then $R(\visited^{(1)}, z)$ is disjoint from $O$\,, and all of the probe walks of $(\visited^{(1)}, z)$
			are strictly shorter than those of $(\visited, v)$\,. Let $W^{(1)} = W v z$\,: because $W$ is a canonical
			walk, $z \in V(W)$ only if $z$ occurs only once in $W$ and is an entry point of $W$ into $\mF$\,,
			in which case the edge $vz$ is never traversed by $W$\,. Then, it is easy to show that $W^{(1)}$
			is a canonical walk for $(\visited^{(1)}, z)$\,. By the inductive hypothesis, \AugmentSearch\ will
			halt on input $(\visited^{(1)}, z)$ and return a value $(\mF, \visited^{(2)}, \fail)$\,, where
			$\visited^{(2)}$ differs from $\visited^{(1)}$ only in that $\visited^{(2)}(x) = \iter$ for all
			$x \in R(\visited^{(1)}, z) \subset R(\visited, v)$\,, and where $W^{(1)}$ is a canonical walk
			for $(\visited^{(2)}, z)$\,. Then, $W$ is a canonical walk for $(\visited^{(2)}, v)$\,.

			Otherwise, if $\visited^{(1)}(z) = \iter$\,, if $v \in I$\,, or if $\mF$ does not cover $v$\,,
			let $\visited^{(2)} = \visited^{(1)}$\,; $W$ is a canonical walk for $(\visited^{(2)}, v)$
			in this case as well.

		\item
			Suppose that at some iteration of the for loop starting at line~\ref{line:augmentSearch-forLoop},
			the data of \AugmentSearch\ is a data pair $(\visited^{(h)}, v)$ for which $W$ is a canonical
			walk,  $\visited^{(h)}(v) = \iter$\,, and $w$ is a neighbor of $v$ satisfying the conditions of 
			lines~\ref{line:augmentSearch-condIter} and~\ref{line:augmentSearch-condOther}. Note that
			$(\visited^{(h)}, w)$ has probe walks ending in $O$ only if $(\visited, v)$ does; then,
			$R(\visited^{(h)}, w)$ is disjoint from $O$\,, and all of the probe walks of the former
			are strictly shorter than those of the latter. Let $W^{(h)} = W v w$\,; because $w$ is not
			covered by $\mF$\,, the fact that $\visited^{(h)}(w) = \visited(w) < \iter$ implies that $w$
			does not occur in $W$\,. Because $w \notin I$, we know that $W^{(h)}$ is a proper alternating
			walk. In particular, $W^{(h)}$ is a canonical walk for $(\visited^{(h)}, w)$\,.

			By the inductive hypothesis, \AugmentSearch\ will then halt on input $(\visited^{(h)}, w)$
			and return a value $(\mF, \visited^{(h+1)}, \fail)$\,, where $\visited^{(h+1)}$ differs from
			$\visited^{(h)}$ only in that $\visited^{(h+1)}(x) = \iter$ for all $x \in R(\visited^{(h)}, w)
			\subset R(\visited, v)$\,, and where $W^{(h)}$ is a canonical walk for $(\visited^{(h+1)}, w)$.
			Then, $W$ is a canonical walk for $(\visited^{(h+1)}, v)$\,.

		\item
			Suppose that at some iteration of the for loop starting at line~\ref{line:augmentSearch-forLoop},
			the data of \AugmentSearch\ is a data pair $(\visited^{(h)}, v)$ for which $W$ is a canonical
			walk,  $\visited^{(h)}(v) = \iter$\,, and $w$ is a neighbor of $v$ satisfying the conditions of 
			lines~\ref{line:augmentSearch-condIter} and~\ref{line:augmentSearch-condEntryPoint}. Then, $w$
			is covered by a path $P \in \mF$ and has a well-defined predecessor $z$ in $P$\,. Let $W^{(h)}
			= W v w z$\,: this is an alternating walk with respect to $\mF$\,.

			The walk $W^{(h)}$ is monotonic only if $w$ is further from the initial point of the path $P \in \mF$
			than any entry point of $W$ on $P$\,. If $P$ contains no entry points of $W$ into $\mF$\,, this is
			satisfied. Otherwise, let $y$ be the final entry point of $W$ into $P$\,.
			\begin{itemize}
			\item
				Suppose that $v$ is not covered by $P$\,. Because $(\visited^{(h)}, v)$ is canonical,
				every	vertex $x$ on the path $P$ from the initial point up to (but possibly not including) $y$
				has $\visited^{(h)}(x) = \iter$\,. Because $\visited^{(h)}(z) < \iter$\,, $z$ is at least
				as far along $P$ as $y$ is; then, $w$ is strictly further. Thus, $W^{(h)}$ is monotonic.

			\item
				If $v$ is covered by $P$\,, then every vertex $x$ on $P$ with $\visited(x) < \iter$ either
				is at least as far as $y$ on $P$\,, or has the property that $y$ is the only entry point between
				$x$ and the end of $P$\,, and that $v$ lies between $x$ and $y$\,. However, if there are more
				than zero vertices of the second type, then $v$ has a predecessor $z$ in $P$ with $\visited(z)
				< \iter$\,. Then, from the analysis of part~\emph{1} above, all vertices $x$ which precede $v$
				in $P$ with $\visited(x) < \iter$ are in $R(\visited^{(1)}, z)$\,, and thus have
				$\visited^{(h)}(x) = \visited^{(2)}(x) = \iter$\,. Then, $W^{(h)}$ is monotonic if and only
				if $w$ is further along $P$ than $y$\,, which reduces to the analysis of the preceding case.
			\end{itemize}
			Because $\visited^{(h)}(z) < \iter$\,, either $z$ is not in $W$\,, or it occurs exactly once
			as an entry point of $W$ into $\mF$\,. Because $\visited^{(h)}(w) < \iter$ and $w$ is further
			along $P$ than any entry point of $W$\,, $w$ does not occur in $W$ at all. Then, neither $vw$
			nor $wz$ are traversed by $W$\,, in which case $W^{(h)}$ is a proper alternating walk. In particular,
			it is a canonical walk for $(\visited^{(h)}, z)$\,.

			Again, $(\visited^{(h)}, z)$ has probe walks ending in $O$ only if $(\visited, v)$ does; then,
			$R(\visited^{(h)}, z)$ is disjoint from $O$\,, and all of the probe walks of the former are
			strictly shorter than those of the latter. By the inductive hypothesis, \AugmentSearch\ will
			then halt on input $(\visited^{(h)}, z)$ and return a value $(\mF, \visited^{(h+1)}, \fail)$\,,
			where $\visited^{(h+1)}$ differs from $\visited^{(h)}$ only in that $\visited^{(h+1)}(x) = \iter$
			for all $x \in R(\visited^{(h)}, z) \subset R(\visited, v)$\,, and where $W^{(h)}$ is a canonical
			walk for $(\visited^{(h+1)}, z)$\,. Then, $W$ is a canonical walk for $(\visited^{(h+1)}, v)$.
		\end{enumerate}
		By induction on the number of neighbors $w \sim v$ satisfying the conditions of
		lines~\ref{line:augmentSearch-condIter}, \ref{line:augmentSearch-condOther},
		and~\ref{line:augmentSearch-condEntryPoint}, the data $(\bar\visited, v)$ when
		the \algorithmicfor\ loop terminates and line~\ref{line:augmentSearch-fail} is
		executed will be a nearly canonical pair, and $\bar\visited$ differs from $\visited$
		only on elements of $R(\visited, v)$\,. 

		It remains to show that $\bar\visited(x) = \iter$ for all $x \in R(\visited, v)$\,. We have shown
		this already for $x = v$\,; then, let $r \in R(\visited, v) \setminus \brace{v}$\,. By definition
		there is a probe walk $\sW$ for $(\visited, v)$ ending in $r$\,. The vertex $w$ immediately
		following $v$ on $\sW$ will be either tested on line~\ref{line:augmentSearch-condReverse} or
		line~\ref{line:augmentSearch-condIter} as a neighbor of $v$\,; then, there exists indices $h'$ such
		that $\sW$ is not a probe walk of $(\visited^{(h')}, v)$\,. Let $h > 0$ be the largest integer
		such that $\sW$ \emph{is} a probe walk for $(\visited^{(h)}, v)$\,: then, there are vertices
		$x \ne v$ in $\sW$ such that $\visited^{(h+1)}(x) = \iter$\,. Let $y \in R$ be the last such
		vertex in $\sW$\,, let $\sW'$ be the segment of $\sW$ from $y$ onwards: then $r \in
		R(\visited^{(h+1)}, y)$\,. Because $\visited^{(h+1)}(y) = \iter$\,, there must have
		been a descendant invocation for $(\visited^{(h)}, v)$ which had input data
		$(\visited^\ast,y)$ for some function $\visited^\ast$\,: it is not difficult to show
		that $\visited^\ast(y) < \iter$\,. By induction on daughter invocations using the analysis
		above, we may show that $(\visited^\ast, y)$ is a canonical data pair with probe walks strictly
		shorter than $\ell$\,: then, for all $x \in R(\visited^\ast, y)$\,, we have $\visited^{(h+1)}(x)
		= \iter$\,. However, because $\visited^{(h+1)}(x) < \iter$ implies $\visited^\ast(x) < \iter$\,,
		and because all vertices $x \in V(\sW)$ after $y$ have $\visited^{(h+1)}(x) < \iter$\,,
		$\sW'$ is a probe walk for $(\visited^\ast, y)$\,. Then, we have $\visited^{(h+1)}(r)
		= \iter$\,.

		Because $\bar\visited(x) = \iter$ if and only if $\visited^{(h)}(x) = \iter$ for some $h \ge 1$\,,
		we then have $\bar\visited(r) = \iter$ for any $r \in R(\visited, v)$\,. By induction, the Lemma
		then follows.
	\end{proof}

	\begin{lemma}
		\label{sublemma:augmentSuccess}
		Suppose that $(\visited, v)$ is a canonical data pair. If $R(\visited, v)$ contains an output vertex,
		\AugmentSearch\ halts on input data $(\visited, v)$\,, with output value $(\mF \oplus \bar\sW,
		\bar\visited, \success)$\,; where $\bar\sW$ is a probe walk for $(\visited, v)$ ending in $O$\,,
		and $\bar\visited$ differs from $\visited$ only in that $\bar\visited(x) = \iter$ only for $x$ in
		some subset of $R(\visited, v)$\,.
	\end{lemma}
	\begin{proof}
		We induct on the length $\ell \in \N$ of the longest probe walk for $(\visited, v)$ ending in $O$\,.
		If $\ell = 0$\,, then $v \in O$\,, and the result holds trivially. Otherwise, suppose $\ell > 0$\
		and that the result holds for those canonical data pairs $(\visited^\ast, x)$ which have probe walks
		of length less than $\ell$ ending in $O$\,.

		Let $W$ be a canonical walk for $(\visited, v)$\,.
		Consider the sequence of vertices $w_1\,, w_2\,, \cdots\,, w_M$ which are tested (either on
		line~\ref{line:augmentSearch-visitReverse}, line~\ref{line:augmentSearch-visitOther}, or 
		line~\ref{line:augmentSearch-visitEntryPoint}) in the course of the invocation of \AugmentSearch.
		We let $\visited^{(1)}$ differ from $\visited$ in that $\visited^{(1)}(v) = \iter$\,,
		and from this define $\visited^{(j)}$ for $j > 1$ by letting $\visited^{(j+1)}$ be the
		second component of the output of the daughter invocation with input data $(\visited^{(j)}, w_j)$\,.
		(If the daughter invocation with data pair $(\visited^{(M)}, w_M)$ halts, this sequence extends to
		$\visited^{(M+1)}$\,.)

		Let $\sW$ be a probe walk for $(\visited, v)$ which ends at a vertex $\omega \in O$\,, and
		let $1 \le N \le M + 1$ be the largest integer such that $R(\visited^{(j)}, w_j)$ is disjoint from $O$
		for all $j < N$\,. If $N = M+1$\,, this means that the invocation of \AugmentSearch\ on input data
		$(\visited^{(M)}, w_M)$ halted with \fail\ in the final part of its' output value, and that there are no
		neighbors of $w \sim v$ which can satisfy the conditions of lines~\ref{line:augmentSearch-condIter},
		\ref{line:augmentSearch-condOther}, and~\ref{line:augmentSearch-condEntryPoint} (due to the choice of $M$
		as the length of the sequence of daughter-invocations).
		However, we may show by induction that for all $1 \le j \le N$\,, $\sW$ is
		a probe walk for $(\visited^{(j)}, v)$\,, which is canonical:
		\begin{itemize}
		\item
			This follows immediately for $j = 1$\,, because $\visited^{(1)}$ only differs from $\visited$ at $v$\,,
			and thus has $\sW$ as a probe walk and $W$ as a canonical walk.
		\item
			Suppose for some $1 \le j < M$ that $W$ is a canonical walk for $(\visited^{(j)}, v)$\,, that
			$\sW$ is a probe walk for $(\visited^{(j)}, v)$\,, and that $\visited^{(j)}(v) = \iter$\,. Then
			we can extend $W$ to a canonical walk $W^{(j)}$ for $(\visited^{(j)}, w_j)$\,: either by setting
			$W^{(j)} = W v w_j$ in the case that $(w_j \arc v) \in A(\mF)$ or $w_j$ is not covered by $\mF$\,,
			or by setting $W^{(j)} = W v z w_j$ where $(z \arc w_j) \in A(\mF)$ otherwise. 
			Because $R(\visited^{(j)}, w_j)$ contains no output vertices, by Lemma~\ref{sublemma:augmentFail}
			we know that $\visited^{(j+1)}$ differs from $\visited^{(j)}$ only on $R(\visited^{(j)}, w_j)$
			and that $W^{(j)}$ is a canonical walk for $(\visited^{(j+1)}, w_j)$\,. Then, $W$ is a canonical
			walk for $(\visited^{(j+1)}, v)$\,.

			For any vertex $x$ in $\sW$\,, the sub-path $\sW_x$ from $x$ to $\omega$ is a probe walk for
			$(\visited^{(j)}, x)$ of length less than $\ell$\,. If $\sW$ has a non-trivial intersection with
			$R(\visited^{(j)}, w_j)$\,, then some vertex $x \in V(\sW)$ is the first such vertex which is given
			as part of an input data pair $(\visited^\ast, x)$ for a descendant invocation for
			$(\visited^{(j)}, w_j)$\,. By induction on the recursion depth from $v$ to $x$\,, we may show
			that there is then a probe walk $\sW^\ast$ for $(\visited^{(j)}, v)$ ending at $x$\,, and
			that $W \sW^\ast$ is a canonical walk for $(\visited^\ast, x)$\,; and precisely because $x$
			is the first vertex of $\sW$ which is visited in a descendant invocation for $(\visited^{(j)}, w_j)$\,,
			we know that  $\visited^\ast(y) < \iter$ for all $y \in V(\sW) \setminus \brace{v,x}$\,. Then, $\sW_x$
			is a probe walk for $(\visited^\ast, x)$\,, and by the induction hypothesis, this invocation of
			\AugmentSearch\ then terminates with \success\ as the last part of its' output. Again by induction
			on the recursion depth, we may also show that the invocation of \AugmentSearch\ with data
			$(\visited^{(j)}, w_j)$ would also terminate with \success\ as the last part of its' output.
			But because $j < N$\,, this cannot happen by Lemma~\ref{sublemma:augmentFail} --- from which
			it follows that $\sW$ is disjoint from $R(\visited^{(j)}, w_j)$\,. Thus $\sW$ is also a probe
			walk for $(\visited^{(j+1)}, v)$\,.
		\end{itemize}
		By induction, $\sW$ is a probe walk for $(\visited^{(N)}, v)$\,, so it must be that $N \le M$\,.
		By the choice of $N$\,, there is then a probe walk $\sW'$ for $(\visited^{(N)}, w_N)$ which
		ends in $O$\,.

		Because $w_N$ is part of the input to a daughter invocation of \AugmentSearch, we have $\visited^{(N)}(w_N)
		< \iter$\,; thus we can easily extend $\sW'$ (by one or two vertices, depending on whether
		$w_N$ is a neighbor of $v$ or the predecessor in $\mF$ of a neighbor of $v$) to form a probe walk $\sW''$
		for $(\visited^{(N)}, v)$\,. Because $\sW''$ will also be a probe walk for $(\visited, v)$\,,
		it has length at most $\ell$\,; then $\sW'$ is strictly shorter than $\ell$ in length. By the
		induction hypothesis, the invocation of \AugmentSearch\ on input data $(\visited^{(N)}, w_N)$ then halts,
		and returns the output value $(\mF \oplus \bar\sW', \bar\visited, \success)$\,, where $\bar\sW'$ is a probe
		walk for $(\visited^{(N)}, w_N)$ ending in $O$\,, and where $\bar\visited$ differs from $\visited^{(N)}$ only
		on a subset of $R(\visited^{(N)}, w_N)$\,. We proceed by cases:
		\begin{itemize}
		\item
			If $v$ is covered by a path of $\mF$\,, $v \notin I$\,, and $w_N$ is the predecessor of $v$ in
			$\mF$\,, then $\bar\sW = v w_N \bar\sW'$ is a probe walk for $(\visited, v)$ ending in $O$\,.
			Note that $A(\mF \oplus \bar\sW) = A(\mF \oplus \bar\sW') \setminus \brace{w_N \arc v}$\,;
			then, the value which is returned as output on line~\ref{line:augmentSearch-successReverse} is
			$(\mF \oplus \bar\sW, \bar\visited, \success)$\,.

		\item
			If $w_N$ is not covered by a path of $\mF$\,, then $w_N \sim v$\,, and the walk $\bar\sW = v w_N
			\bar\sW'$ is a probe walk for $(\visited, v)$ ending in $O$\,. Note that $A(\mF \oplus \bar\sW)
			= A(\mF \oplus \bar\sW') \union \brace{v \arc w_N}$\,; then, the value which is returned as output
			on line~\ref{line:augmentSearch-successOther} is $(\mF \oplus \bar\sW, \bar\visited, \success)$\,.
			
		\item
			If neither of the previous two cases apply, it must be that $w_N$ is the predecessor in $\mF$ of some third vertex
			$u \sim v$\,. Because $w_N$ is part of the input to a daughter invocation for $(\visited^{(N)}, v)$\,,
			we know that $\visited^{(N)}(u) < \iter$\,: then, $\bar\sW = v u\!\: w_N \bar\sW'$ is a probe walk for
			$(\visited, v)$ ending in $O$\,. Note that $A(\mF \oplus \bar\sW) = \sqparen{A(\mF \oplus \bar\sW)
			\setminus \brace{w_N \arc u}} \union \brace{v \arc u}$\,; then, the value which is returned as output
			on line~\ref{line:augmentSearch-successEntryPoint} is $(\mF \oplus \bar\sW,\, \bar\visited,\, \success)$\,.
		\end{itemize}
		Finally, because $R(\visited^{(N)}, w_N) \subset R(\visited, v)$\,, and because $\visited^{(N)}$ differs from
		$\visited$ only on $R(\visited{(i)}, w_i) \subset R(\visited, v)$ for $1 \le i < N$\,, it follows that
		$\bar\visited$ differs from $\visited$ only on a subset of $R(\visited, v)$\,, with $\bar\visited(x) = \iter$
		on that subset. Thus, if the Lemma holds for pairs $(\visited, v)$ having probe walks of length less than
		$\ell \ge 0$ ending in $O$\,, it also holds for such pairs with probe walks ending in $O$ of length $\ell + 1$\,.
		By induction, the Lemma then holds.
	\end{proof}

	To prove the Theorem, it then suffices to note that for a function $\visited: V(G) \to \N$ with $\visited(x) < \iter$
	for all $x \in V(G)$\,, probe walks for $(\visited, i)$ are just proper alternating walks with respect to $\mF$
	which start at $i$\,, in which case such a probe walk $\bar\sW$ which ends in $O$ is a proper augmenting walk for
	$\mF$\,. Then all the various parts of the Theorem follow from Lemmas~\ref{sublemma:augmentFail}
	and~\ref{sublemma:augmentSuccess} collectively.
	\setcounter{theorem}{\value{tmpTheorem}}
	\renewcommand\qed\oldQed
\end{proof}

\paragraph*{Run-time analysis.}

Because \AugmentSearch\ marks each vertex $v$ with $\visited(v) \gets \iter$ when it visits $v$\,,
each vertex is only visited once. At each vertex, each of the neighbors $w \sim v$ are tested for if
they fulfill the condition of line~\ref{line:augmentSearch-condReverse}, or of lines~\ref{line:augmentSearch-condIter},
\ref{line:augmentSearch-condOther}, and~\ref{line:augmentSearch-condEntryPoint}. Because computing
$\prev$\,, $\AddArc$\,, and $\RemoveArc$ is constant-time for $\mF$ a collection of vertex-disjoint
paths (or differing only slightly from one as described in the discussion on implementation details),
the amount of work in an invocation to \AugmentSearch\ for a vertex $v \in V(G)$ is $O(\deg v)$\,,
neglecting the work performed in descendant invocations. 
Summing over all vertices $v \in V(G)$\,, the run-time of \AugmentSearch\ is then $O(m)$
for an input as described in the statement of Theorem~\ref{thm:augmentSearch}.

\subsubsection{An efficient algorithm for constructing a path cover for $(G,I,O)$}

Using \AugmentSearch\ as a subroutine to build successively larger proper families of vertex-disjoint
$I$ -- $O$ paths, Algorithm~\ref{alg:buildPathFamily} describes a straightforward subroutine which
attempts to build a path cover for $(G,I,O)$\,. 

\begin{algorithm}[ht]
	\mycaption
		{\BuildPathCover(G, I, O)}
		{tries to build path cover for $(G,I,O)$}
	\label{alg:buildPathFamily}
	\begin{algorithmic}[1]
	\REQUIRE $(G,I,O)$ is a geometry.

	\medskip

	\setlength\commentskip{52ex}
	\LET	$\mF$: an empty collection of vertex-disjoint dipaths in $G$

	\smallskip

	\LET	$\visited: V(G) \to \N$ be an array initially set to zero
	\LET	$\iter \gets 0$
	\FORALL {$i \in I$}
		\label{line:buildPathCover-forLoop}
		\STATE $\iter \gets \iter + 1$
		\STATE $(\mF, \visited, \status) \gets \AugmentSearch(G,I,O,\mF,\iter,\visited,i)$
		\IFTHEN {$\status = \fail$} \algorithmicreturn\ \fail
			\label{line:buildPathCover-testAugment}
	\ENDFOR

	\smallskip

	\IF {$V(G) \setminus V(\mF) = \vide$}
		\RETURN $\mF$
	\ELSE
		\RETURN \fail
			\label{line:buildPathCover-testCover}
	\ENDIF	
	\end{algorithmic}
\end{algorithm}

\begin{corollary}
	\label{cor:buildPathCover}
	Let $(G,I,O)$ be a geometry with $\abs{I} = \abs{O}$\,: then \BuildPathCover\ halts on input
	$(G,I,O)$\,. Furthermore, let $\sigma = \BuildPathCover(G,I,O)$\,. If $\sigma = \fail$\,, then
	$(G,I,O)$ does not have a causal flow; otherwise, $\sigma$ is a path cover $\mF$ for $(G,I,O)$\,.
\end{corollary}
\begin{proof}
	Suppose $(G,I,O)$ has a causal flow: then it has a collection of $k = \abs{I} = \abs{O}$
	vertex-disjoint $I$ -- $O$ paths by Lemma~\ref{thm:orbitPathCover}. Then, by 
	Corollary~\ref{cor:freeProperAugmentingWalk}, for any proper collection $\mF$ of
	vertex-disjoint $I$ -- $O$ paths with $\abs{\mF} < k$\,, there is a proper augmenting
	walk for $\mF$ starting at any $i \in I$ which is not covered by $\mF$\,. For such a
	collection $\mF$ and vertex $i$\,, if $\visited(x) < \iter$ for all $x \in V(G)$\,,
	$\AugmentSearch(G,I,O,\mF,\iter,\visited,i)$ returns $(\mF \oplus W, \bar\visited, \success)$\,,
	where $\bar\visited(x) \le \iter$ for all $x \in V(G)$\,, and where $W$ is a proper
	augmenting walk for $\mF$ starting at $i$\,. Then, $\mF \oplus W$ is a proper
	collection of vertex-disjoint paths, covering $i$ and the input vertices covered by
	$\mF$\,, and with $\abs{\mF \oplus W} = \abs{\mF} + 1$\,. By induction, we may then show
	that at the end of the \algorithmicfor\ loop starting at line~\ref{line:buildPathCover-forLoop},
	$\mF$ will be a family of vertex-disjoint $I$ -- $O$ paths which covers all of $I$\,,
	in which case $\abs{\mF} = k$\,. If all of the vertices of $V(G)$ are covered by $\mF$\,,
	$\mF$ is then a path cover for $(G,I,O)$\,, and \BuildPathCover\ returns $\mF$\,.
	Taking the contrapositive, if $\BuildPathCover(G,I,O)$ returns \fail, then $(G,I,O)$ has no path
	cover.

	Conversely, if $\BuildPathCover(G,I,O)$ returns \fail, then either the condition
	of line~\ref{line:buildPathCover-testAugment} failed, or the condition of
	line~\ref{line:buildPathCover-testCover} failed. If the former is true,
	then by Theorem~\ref{thm:augmentSearch} there were no proper augmenting walks for
	some proper collection $\mF$ of fewer than $k$ disjoint $I$--$O$ paths,
	in which case by Corollary~\ref{cor:freeProperAugmentingWalk} there is no such
	collection of size $k$\,, and thus no causal path cover for $(G,I,O)$\,. Otherwise,
	$\mF$ is a maximum-size collection of disjoint paths from $I$ to $O$\,, but is not a path
	cover for $(G,I,O)$\,; then by Theorem~\ref{thm:uniqueFamilyPaths}, there again is no
	causal path cover for $(G,I,O)$\,. In either case, there is no causal flow for
	$(G,I,O)$ by Theorem~\ref{thm:graphth-charn}. The result then holds.
\end{proof}

\paragraph{Run-time analysis.}

\BuildPathCover\ iterates through $k = \abs{I}$ input vertices as it increases the size of
the collection of vertex-disjoint paths, invoking \AugmentSearch\ for each one. The running
time for this portion of the algorithm is then $O(km)$\,. As this is larger than the time
required to initialize $\visited$ or to determine if there is an element $v \in V(G)$
such that $v \notin V(\mF)$\,, this dominates the asymptotic running time of \BuildPathCover.

\subsection{Efficiently finding a causal order for a given successor function}
\label{sec:findCausalOrder}

Given a path cover $\mC$ for a geometry $(G,I,O)$\,, and in particular the successor
function $f$ of $\mC$\,, we are interested in determining if the natural pre-order $\preceq$
for $f$ is a partial order, and constructing it if so. 
In this section, I present an efficient algorithm to
determine whether or not $\preceq$ is a partial order, by reduction
to the transitive closure problem on digraphs. 

\subsubsection{The Transitive Closure Problem}

Any binary relation $\textsf{R}$ can be regarded as defining a digraph $D$ with $(x \arc y) \in A(D)
\iff (x \textsf{R} y)$\,. Chains of related elements can then be described by directed walks in the
digraph $D$\,. This motivates the following definition:

\begin{definition}
	\label{dfn:influencingDigraph}
	Let $f$ be a successor function for a geometry $(G,I,O)$\,: the \emph{influencing digraph}
	$\sI_f$ is then the directed graph with vertices $V(\sI_f) = V(G)$\,, where 
	$(x \arc y) \in A(\sI_f)$ if one of $y = x$\,, $y = f(x)$\,, or $y \sim f(x)$ hold.
\end{definition}

The three types of arcs in Definition~\ref{dfn:influencingDigraph} correspond to the relations in
Equation~\ref{eqn:naturalPreorder}, whose transitive closure is the natural pre-order. Note that aside
from self-loops $x \arc x$\,, the arcs in $\sI_f$ correspond directly to the two varieties of segments
of influencing walks. (This is an alternative way of proving Lemma~\ref{thm:influencingNatlPreorder}.)

\vspace{1ex}
\begin{figure}[ht]
	\setlength\unitlength{0.16mm}
	\begin{center}
		\begin{minipage}[c]{45mm}
 		\begin{picture}(256,175)
 			\includegraphics[width=41mm]{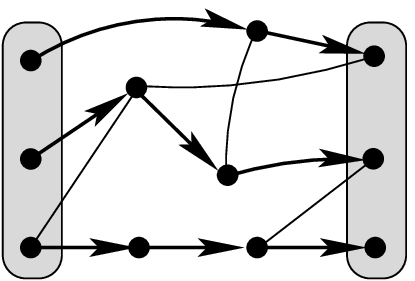}
			\puttext(-261,90)[r]{$I$}
			\puttext(5,90)[l]{$O$}
 		\end{picture}
		\end{minipage}
		$\qquad\mapsto\qquad\quad$
		\begin{minipage}[c]{45mm}
 		\begin{picture}(256,175)
 			\includegraphics[width=41mm]{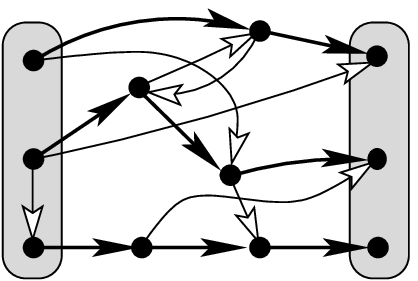}
			\puttext(-261,90)[r]{$I$}
			\puttext(5,90)[l]{$O$}
 		\end{picture}
		\end{minipage}
		\iftwocol{\caption}{\figcaption}
 					{On the left: a geometry $(G,I,O)$ with a path cover $\mC$\,. Arrows represent the action of the successor function
					$f: O\comp \to I\comp$ of $\mC$\,. On the right: the corresponding influencing digraph $\sI_f$\,. Solid arrows
					represent arcs of the form $x \arc f(x)$\,, and hollow arrows represent arcs $x \arc y$ for $y \sim f(x)$\,.
					(Self-loops are omitted for clarity.)}
 		\label{fig:inflDigraph}
	\end{center}
\end{figure}
\vspace{-1ex}

It is natural to also speak of transitive
closures of binary relations in graph-theoretic terms, as follows:

\begin{definition}
	For a digraph $D$\,, the \emph{transitive closure} of $D$ is the digraph $T$ with
	$V(T) = V(D)$\,, and such that $(x \arc y) \in A(T)$ if and only if there is a
	non-trivial\footnote{Note that if a vertex $x$ has a loop $x \arc x$ (which are permitted
	in digraphs), then the directed walk $x \arc x$ is a non-trivial walk.} directed walk
	from $x$ to $y$ in the digraph $D$\,.
\end{definition}

Thus, $x \arc y$ is an arc in the transitive closure of $\sI_f$ and only if $x \preceq y$\,, or
equivalently iff there is an influencing walk for $\mC$ from $x$ to $y$ in $G$.

\begin{theoremname}[Transitive Closure Problem]
	Given a digraph $D$\,, determine it's transitive closure $T$\,.
\end{theoremname}

The Transitive Closure problem is known to be efficiently solvable. Algorithm~\ref{alg:fig3-8} presents
on solution, which is (a paraphrasing of) the pseudocode of Figure~3.8 from~\cite{N95}.
This algorithm is a simple modification of Tarjan's algorithm for finding \emph{strongly connected
components} of digraphs (equivalence classes of mutually reachable vertices using directed walks),
which finds the transitive closure by determining the ``descendants'' of each $x \in V(D)$\,:
\begin{align}
	\label{eqn:succDfn}
		\Desc(x)
	\;=\;
		\set{\, y \in V(D)}{\big. \text{$D$ contains a non-trivial directed walk from $x$
		to $y$}}	\;\;.
\end{align}
\vspace{-2ex}

\begin{algorithm}[ht]
	\mycaption
		{\textrm{Figure~3.8 of~\cite{N95}}}
		{an algorithm for transitive closure of a digraph}
	\label{alg:fig3-8}
	\begin{algorithmic}[1]
	\PROCEDURE \SimpleTC(v)
	\BEGIN
		\STATE	$\Root(v) \gets v\;;\; \Comp(v) \gets \nil$
			\label{line:setNilComponent}
		\STATE	$\PUSH(v, \stack)$
		\STATE	$\Desc(v) \gets \set{w \in V(D)}{(v \arc w) \in A(D)}$
			\label{line:initDesc}
		\FORALL {$\fwdarc{v}{w}{D}$}
			\label{line:fwdArc}
			\IFTHEN {($w$ is not already visited)}	$\SimpleTC(w)$
				\label{line:visitedDependant}
			\IFTHEN {$\Comp(w) = \nil$}				$\Root(v) \gets \min (\Root(v), \Root(w))$
				\label{line:updateRoot}
			\STATE	$\Desc(v) \gets \Desc(v) \union \Desc(w)$
				\label{line:augmentDesc}
		\ENDFOR
		\IF {$\Root(v) = v$}
			\label{line:SimpleTC-superfluous-begin}
			\STATE	create a new component $C$
			\REPEAT
				\LET	$w \gets \POP(\stack)$
					\label{line:popStackDetermCompon}
				\STATE	$\Comp(w) \gets C$
				\STATE	insert $w$ into the component $C$
				\STATE	$\Desc(w) \gets \Desc(v)$
			\UNTIL {$w = v$}
		\ENDIF
		\label{line:SimpleTC-superfluous-end}
	\END
	
	\medskip

	\PROCEDURE \texttt{main}
	\BEGIN 
		\LET	$\stack \gets \vide$
		\FORALL {$v \in V(D)$}
			\IFTHEN {($v$ is not already visited)}		$\SimpleTC(v)$
		\ENDFOR
	\END
	\end{algorithmic}
\end{algorithm}

The following is an overview of Algorithm~\ref{alg:fig3-8}: interested readers may refer
to \cite{N95} for a more complete analysis.
\begin{itemize}
\item
	\label{discn:TarjanDepthFirst}
	A di-connected component of $D$ is an equivalence class of vertices which can be reached
	from each other by non-trivial directed walks in $D$\,. Tarjan's algorithm detects these
	components by performing a depth-first search which traverses arcs of $D$\,, and detecting
	when it has traversed a directed cycle in $D$\,.

\item
	\label{discn:TarjanStackTracking}
	A stack is used to keep track of vertices of the digraph have been visited, but
	whose di-connected component has not yet been completely determined. When the vertices
	belonging to a a given component are determined, we pop them off of the stack
	(line~\ref{line:popStackDetermCompon}) and insert them into a set representing
	that component.

\item
	\label{discn:TarjanComponentRoot}
	We say that $v$ precedes $w$ in the ordering of the stack if $v$ is on the
	stack and $w$ is not, or if $v$ is lower on the stack than $w$ is. Then, we may
	keep track of the ``root'' $\Root(v)$ of $v$\,, which is an upper bound on the
	stack-minimal vertex of the component containing $v$. At first, we set the root of $v$ to
	itself, and we always ensure that $\Root(v) \le v$\,.
	
	Suppose we discover a descendant $w$ of $v$ such that $\Root(w) \le \Root(v) \le v$\,.
	Then $v$ is a descendant of $\Root(w)$\,, which is in a common component with $w$ by
	definition. Because $w$ is also a descendent of $v$\,, $v$ must be in a common
	component with $w$\,. Then $\Root(w)$ is the smallest known vertex in that
	component: we update $\Root(v) \gets \Root(w)$ to improve the known minimum for $v$\,.

\item
	\label{discn:TarjanUpdateRoot}
	Because vertices are only allocated to a di-connected component after they are popped
	off the stack, we may test each of the descendants $w$ of $v$ to see if they have been
	allocated to a component, rather than testing if $\Root(w) \le \Root(v)$\,. If not,
	then $v$ is in a common component with $w$\,, and we update $\Root(v)$ to be the minimum
	of $\Root(v)$ and $\Root(w)$ on line~\ref{line:updateRoot}, as in the previous case.

\item
	\label{discn:TarjanComponentAllocation}
	If $\Root(v) = v$ on line~\ref{line:SimpleTC-superfluous-begin}, then $v$ is the
	stack-minimal element of its' component: then any vertices higher than $v$ on the stack will
	be in the same component as $v$\,. Conversely, because all descendants of $v$ have been
	visited by that point, all of the vertices in the same component as $v$ are still on the
	stack. Thus, we may pop them off the stack and allocate them to a component, until we have
	removed $v$ off of the stack (lines~\ref{line:SimpleTC-superfluous-begin}
	through~\ref{line:SimpleTC-superfluous-end}).

\item
	\label{discn:TarjanDescendantOrdering}
	As we determine the connected components of the digraph, we may maintain the sets of
	descendants of each vertex: if $(v \arc w) \in A(D)$\,, then the descendants of $w$ are
	all also descendants of $v$\,, so we ensure that $\Desc(w) \subset \Desc(v)$
	(as on line~\ref{line:augmentDesc}).
\end{itemize}
The above is performed for all vertices $v \in V(G)$ to obtain the transitive closure.

Algorithm~\ref{alg:fig3-8} is sufficient to build the natural pre-order $\preceq$ for a successor
function $f$\,. However, the output does not indicate whether $\preceq$ is a partial order, and it
performs work that is unnecessary if $\preceq$ is not actually a partial order. We may also take
advantage of the availability of the path cover $\mC$ which is given as input, which is not available
in the more general Transitive Closure problem. Therefore, we are interested in adapting 
Algorithm~\ref{alg:fig3-8} to the application of finding a causal order.

\subsubsection{Chain decompositions with respect to the path cover $\mC$}
\label{sec:chainDecomp}

Let $\mC$ be a path cover for $(G,I,O)$ with successor function $f$\,. The transitive closure 
of the influencing digraph $\sI_f$ will often have high maximum degree: because the longest path in
$\mC$ has at least $n/k$ vertices, and the end-point of this path will be at the
terminus of arcs coming from every vertex on the path, the maximum in-degree of the transitive
closure is at least $n/k$\,; and similarly for the maximum out-degree. In
Algorithm~\ref{alg:fig3-8}, this implies that the set $\Desc(v)$ may become comparable to
$V(G)$ in size. In order to construct the arc-lists of the transitive closure reasonably efficiently, we
want to reduce the effort required in determining the sets $\Desc(v)$\,.
 
A standard approach to this problem would be to find a \emph{chain decomposition}~\cite{N95} for
$\sI_f$\,, which is a collection of vertex-disjoint dipaths of $\sI_f$ which cover all of $\sI_f$\,.
By the definition of the influencing digraph, $\mC$ itself is such decomposition of $\sI_f$\,.
Then, using a chain decomposition with respect to $\mC$\,, we can efficiently represent
$\Desc(x)$ in terms of the first vertex $y$ in each path of $\mC$ such that $y \in \Desc(x)$\,.

\begin{definition}
	Let $\mC = \brace{\scrP_j}_{j \in K}$ be a parameterization of the paths of a path
	cover $\mC$ for a geometry $(G,I,O)$\,, let $f$ be the successor function of $\mC$\,,
	and let $\preceq$ be the natural pre-order for $f$\,. Then, for	$x \in V(G)$ and
	$j \in K$\,, the \emph{supremum $\sup_j(x)$ of $x$ in $\scrP_j$} is the minimum integer
	$m \in \N$\,, such that $x \preceq y$ for all vertices $y \in V(\scrP_j)$ which are
	further than distance $m$ from the initial vertex of $\scrP_j$\,.
\end{definition}

We may use the suprema of $x$ in the paths of $\mC$ to characterize the natural pre-order
for $f$\,:
\begin{lemma}
	\label{lemma:supremaCharn}
	Let $\mC = \brace{\scrP_j}_{j \in K}$ be a parameterization of the paths of a path
	cover $\mC$ for a geometry $(G,I,O)$\,, let $f$ be the successor function of $\mC$\,,
	let $\preceq$ be the natural pre-order for $f$\,, and let $L: V(G) \to \N$ map 
	vertices $x \in V(G)$ to the distance of $x$ from the initial point of the path
	of $\mC$ which contains $x$\,. Then
	\begin{align}
			x \preceq y
		\quad\iff\quad
			\sup\nolimits_j(x) \le L(y)
	\end{align}
	holds for all $x \in V(G)$ and $y \in V(\scrP_j)$\,, for any $j \in K$\,.
\end{lemma}
\begin{proof}
	Let $x \in V(G)$\,, and fix $\scrP_j \in \mC$\,. Let $v \in V(\scrP_j)$ be such that
	$L(v) = \sup_j(x)$\,. By definition, if $y \in V(\scrP_j)$ and $x \preceq y$\,,
	then $L(y) \ge L(v)$\,. Conversely, if $y \in V(\scrP_j)$ and $L(y) = L(v) + h$ for
	$h \ge 0$\,, then $y = f^h(v)$\,; then $x \preceq v \preceq y$\,, and the result
	holds by transitivity.
\end{proof}


To determine the supremum function for all vertices, it will be helpful to be able to efficiently
determine which path of $\mC$ a given vertex belongs to and how far it is from the initial vertex
for it's path. Algorithm~\ref{alg:getChainDecomp} describes a simple procedure to do this, which
also produces the successor function for the path cover $\mC$\,. (In the case where $\abs{I} =
\abs{O}$\,, every path of $\mC$ has an initial point in $I$\,; we then take $K = I$ to be
the index set of the paths of $\mC$\,.)

\begin{algorithm}[th]
	\mycaption
		{\GetChainDecomp(G, I, O, \mC)}
		{obtain the successor function $f$ of $\mC$\,, and obtain functions describing the chain decomposition
		of the influencing digraph $\sI_f$}
		\label{alg:getChainDecomp}
	\begin{algorithmic}[1]
	\REQUIRE $(G,I,O)$ be a geometry with $\abs{I} = \abs{O}$
	\REQUIRE $\mC$ a path cover of $(G,I,O)$

	\medskip
				
	\LET	$P: V(G) \to I$ an array
	\LET	$L: V(G) \to \N$ an array
	\LET	$f: O\comp \to I\comp$ an array
	\FORALL {$i \in I$}
		\LET $v \gets i$\,,\; $\ell \gets 0$
		\WHILE {$v \notin O$}
			\STATE $f(v) \gets \next(\mC, v)$
			\STATE $P(v) \gets i\,;\; L(v) \gets \ell$
			\STATE $v \gets f(v)$
			\STATE $\ell \gets \ell + 1$
		\ENDWHILE
		\STATE $P(v) \gets i\,;\; L(v) \gets \ell$
	\ENDFOR
	\RETURN $(f, P, L)$
	\end{algorithmic}
\end{algorithm}


\subsubsection{Detecting vicious circuits with respect to $\mC$}

If the influencing digraph $\sI_f$ contains non-trivial di-connected components, we know
that there are closed influencing walks --- i.e. vicious circuits --- for $\mC$ in $(G,I,O)$\,.
In that case, Theorem~\ref{thm:graphth-charn} together with Theorem~\ref{thm:uniqueFamilyPaths}
imply that $(G,I,O)$ has no causal flow, in which case we may as well abort. 
Recall that $\SimpleTC$ keeps track of di-connected components by allocating vertices to
a component $C$ after the elements of $C$ have been completely determined. However, the state of
being allocated into a component can be replaced in this analysis by \emph{any} status of the
vertex which is changed after the descendants of a vertex have been determined; and this status
may be used to determine if a vicious circuit has been found.


Algorithm~\ref{alg:initStatus} is a simple procedure to initialize an array \status\ over $V(G)$\,.
A status of \none\ will indicate that no descendants of the vertex have been determined (except itself),
\fixed\ will indicate that all descendants of the vertex have been determined, and \pending\
will indicate that the descendants are in the course of being determined. Because output vertices
have only themselves for descendants, their status is initialized to \fixed; all other vertices
are initialized with $\status(v) = \none$\,. At the same time, Algorithm~\ref{alg:initStatus}
initializes a supremum function which represents only the relationships of each vertex to the ones
following it on the same path. 
\begin{algorithm}[th]
	\mycaption
		{\InitStatus(G, I, O, P, L)}
		{initialize the supremum function, and the status of each vertex}
		\label{alg:initStatus}
	\begin{algorithmic}[1]
	\REQUIRE $(G,I,O)$ is a geometry
 	\REQUIRE $P: V(G) \to I$ maps each $x \in V(G)$ to $i \in I$ such that $x$ is in the orbit of $i$ under $f$
 	\REQUIRE $L: V(G) \to \N$ maps each $x \in V(G)$ to $h \in \N$ such that $x = f^h(P(x))$
	
	\medskip

	\LET $\Sup: I \x V(G) \to \N$ an array
	\LET $\status: V(G) \to \brace{\none, \pending, \fixed}$ an array

	\FORALL {$v \in V(G)$}
		\FORALL {$i \in I$}
			\IFTHEN {$i = P(v)$} $\Sup(i,v) \gets L(v)$
			\STATE\algorithmicelse\ $\Sup(i,v) \gets \abs{V(G)}$
		\ENDFOR
		\IFTHEN {$v \in O$} $\status(v) \gets \fixed$
		\STATE\algorithmicelse\ $\status(v) \gets \none$
	\ENDFOR
	\RETURN $(\Sup, \status)$
	\end{algorithmic}
\end{algorithm}

\subsubsection{An efficient algorithm for computing the natural pre-order of $f$}

Algorithms~\ref{alg:traverseInflWalk} and~\ref{alg:computeSuprema} below represent a modified
version of Algorithm~\ref{alg:fig3-8}, specialized to the application of computing the natural
pre-order for the successor function $f$ of a path cover $\mC$\,. Rather than explicitly
constructing the influencing digraph $\sI_f$ and traversing directed walks in $\sI_f$ (as is
done in Algorithm~\ref{alg:fig3-8}), we instead traverse influencing walks for $\mC$ (characterized
by its' successor function) in the graph $G$\,.

\begin{algorithm}[th]
	\mycaption
		{\TraverseInflWalk(G, I, O, f, \Sup, \status, v)}
		{compute the suprema of $v$ and all of its' descendants, by traversing influencing walks from $v$}
		\label{alg:traverseInflWalk}
	\begin{algorithmic}[1]
	\REQUIRE $(G,I,O)$ is a geometry
	\REQUIRE $f: O\comp \to I\comp$ is a successor function for $(G,I,O)$
	\REQUIRE $\Sup: I \x V(G) \to \N$
	\REQUIRE $\status: V(G) \to \brace{\none, \pending, \fixed}$ 
	\REQUIRE $v \in O\comp$
	
	\medskip
	\setlength\commentskip{45ex}
	
	\STATE $\status(v) \gets \pending$
		\label{line:setPending}
		\FORALL {$w = f(v)$ \AND\ \algorithmicforall\ $w \sim f(v)$}
			\label{line:forInflSegments}
			\IF {$w \ne v$}
				\IFTHEN {$\status(w) = \none$} $(\Sup, \status) \gets \TraverseInflWalk(G,I,O,f,\Sup,\status,w)$
				\IF {$\status(w) = \pending$}
					\label{line:recurAbortTest}
					\RETURN $(\Sup, \status)$
				\ELSE
					\FORALL {$i \in I$}
						\label{line:improveSuprema}
						\IFTHEN {$\Sup(i, v) > \Sup(i, w)$} $\Sup(i,v) \gets \Sup(i,w)$
					\ENDFOR
				\ENDIF
			\ENDIF
		\ENDFOR
	\STATE $\status(v) \gets \fixed$
		\label{line:fixStatus}
	\RETURN $(\Sup, \status)$
	\end{algorithmic}
\end{algorithm}
\begin{algorithm}[th]
	\mycaption
		{\ComputeSuprema(G,I,O,f,P,L)}
		{obtain the successor function $f$ of $\mC$\,, and compute the natural pre-order of $f$
		in the form of a supremum function and functions characterizing $\mC$}
		\label{alg:computeSuprema}

	\begin{algorithmic}[1]
	\REQUIRE $(G,I,O)$ is a geometry with $\abs{I} = \abs{O}$ and successor function $f: O\comp \to I\comp$
 	\REQUIRE $P: V(G) \to I$ maps each $x \in V(G)$ to $i \in I$ such that $x$ is in the orbit of $i$ under $f$
 	\REQUIRE $L: V(G) \to \N$ maps each $x \in V(G)$ to $h \in \N$ such that $x = f^h(P(x))$

	\medskip

	\LET $(\Sup, \status) \gets \InitStatus(G,I,O,P,L)$
	\FORALL {$v \in O\comp$}
		\label{line:beginTarjan}
		\IFTHEN {$\status(v) = \none$} 		$(\Sup, \status) \gets \TraverseInflWalk(G, I, O, f, \Sup, \status, v)$
		\IFTHEN {$\status(v) = \pending$}	\algorithmicreturn\ $\fail$
				\label{line:abortTest}
	\ENDFOR
		\label{line:endTarjan}
	\RETURN $\Sup$
	\end{algorithmic}
\end{algorithm}

\begin{theorem}
	\label{thm:buildCausalOrder}
	Let $f$ be a successor function of a path cover $\mC$ for a geometry $(G,I,O)$\,.
	Let $P: V(G) \to I$ map vertices $v$ to the initial point of the path of $\mC$ that
	covers $v$\,, and let $L: V(G) \to \N$ map vertices $v$ to the integer $h \in \N$
	such that $v = f^h(P(v))$\,. Then $\ComputeSuprema$ halts on input $(G,I,O,f,P,L)$\,.
	Furthermore, let $\sigma = \ComputeSuprema(G,I,O,f,P,L)$\,. If $\sigma = \fail$\,,
	then $(G,I,O)$ does not have a causal flow; otherwise, $(G,I,O)$ does have a
	causal flow, and $\sigma$ is a supremum function $\Sup: I \x V(G) \to \N$ satisfying
	\begin{align}
		\label{eqn:correctPreorderRelation}
			x \preceq y
		\quad\iff\quad
			\Sup(P(y), x) \le L(y)
	\end{align}
	for all $x, y \in V(G)$\,, where $\preceq$ is the natural pre-order for $f$\,.
\end{theorem}
\begin{proof}
	We will reduce the correctness of Algorithms~\ref{alg:traverseInflWalk} and~\ref{alg:computeSuprema}
	to that of Algorithm~\ref{alg:fig3-8}, where $D = \sI_f$ is the digraph provided as the the input
	of the main procedure. Throughout, $\preceq$ denotes the natural pre-order of $f$\,.

	For distinct vertices $v, w \in V(G)$\,, because $(v \arc w) \in A(\sI_f)$ if and only if either $w = f(v)$
	or $w \sim f(v)$\,, we may replace the iterator limits ``$\fwdarc{v}{w}{D}$'' of the \algorithmicfor\ loop
	starting on line~\ref{line:fwdArc} of Algorithm~\ref{alg:fig3-8} with a loop iterating over $w = f(v)$ and
	$w \sim f(v)$\,: this is what we have on line~\ref{line:forInflSegments} of \TraverseInflWalk.

	At line~\ref{line:updateRoot} of Algorithm~\ref{alg:fig3-8}, if $\Comp(w) = \nil$\,, we
	infer that $v$ and $w$ are in a common di-connected component of the digraph $\sI_f$\,: this
	implies that $v \preceq w$ and $w \preceq v$\,. If $v \ne w$\,, this implies that $\preceq$
	is not antisymmetric, and thus not a partial order; by Theorem~\ref{thm:graphth-charn},
	$\mC$ is then not a causal path cover. We proceed by cases:
	\begin{itemize}
	\item
		If $\preceq$ is antisymmetric, then the influencing digraph is acyclic, in which case
		$\sI_f$ has only trivial di-connected components. In this case, the following changes
		preserve the functionality of Algorithm~\ref{alg:fig3-8}:
		\renewcommand\labelitemii{---}
		\begin{itemize}
		\item
			In the case that $w = v$ in the \algorithmicfor\ loop, all the operations performed
			are superfluous, in which case we may embed lines~\ref{line:visitedDependant}
			through~\ref{line:augmentDesc} in an \algorithmicif\ statement conditioned on
			$w \ne v$\,.
		
		\item
			Because each vertex is the only vertex in its' component when $\sI_f$ is acyclic, we may
			replace lines~\ref{line:SimpleTC-superfluous-begin} through~\ref{line:SimpleTC-superfluous-end}
			of $\SimpleTC$ with a line setting $\Comp(v)$ to an arbitrary non-\nil\ value, which
			in this case may be interpreted as allocating the vertex $v$ to its' di-connected
			component (i.e. the singleton $\brace{v}$). Also, the condition of line~\ref{line:updateRoot}
			is never satisfied in a call to $\SimpleTC(v)$\,. Then, we may replace the conditional
			code with an arbitrary statement, e.g. a command to abort the procedure. 

		\item
			After the above replacement, the value of $\stack$ is not used within the
			procedure call $\SimpleTC(v)$\,, and has the same value after the procedure call
			to $\SimpleTC(v)$ as it does before the call. Then, $\stack$ is superfluous to the
			performance of the algorithm. Similarly, the value of $\Root(v)$ is not affected
			except to initialize it. We may then eliminate all references to either one.

		\item
			The value of $\Comp(w)$ is only tested to determine whether or not it is
			$\nil$\,, so we may replace the array $\Comp$ with $\status$\,, and its' possible
			states of being $\nil$ or non-$\nil$ with the states of being $\pending$ and
			non-$\pending$\,. We define the two values \none\ and \fixed\ to represent
			being non-\pending\ and also having not yet been visited, and being non-\pending\
			and having been visited, respectively.

		\item
			Using the array \Sup\ to implicitly represent the sets of descendants, we may
			replace the union performed on line~\ref{line:augmentDesc} with code which sets
			$\Sup(i,v)$ to the minimum of $\Sup(i,v)$ and $\Sup(i,w)$ for each $i \in I$\,.
			Note also that, because $x$ is a descendant of $v$ in $\sI_f$ iff $v = x$ or
			$x$ is a descendant of $w$\,, we may remove the initialization of $\Desc(v)$ on
			line~\ref{line:initDesc} of Algorithm~\ref{alg:fig3-8} if we initialize $\Sup$
			for each vertex so that it represents each vertex as a descendant of itself
			(for instance, in the main procedure, which is replaced by \ComputeSuprema).
		\end{itemize}
		By performing the substitutions described above, we can easily see that $\TraverseInflWalk$
		together with $\ComputeSuprema$ is equivalent to Algorithm~\ref{alg:fig3-8} when
		$\preceq$ is anti-symmetric. Then, $\sigma \ne \fail$ because line~\ref{line:abortTest}
		of $\ComputeSuprema$ is never evaluated; we then have $\sigma = \Sup$ as in
		Equation~\ref{eqn:correctPreorderRelation}, from the correctness of
		Algorithm~\ref{alg:fig3-8}.

	\item
		If $\preceq$ is not antisymmetric, then there are distinct vertices $x, y \in V(G)$ such
		that $x \preceq y \preceq x$\,, in which case $x$ and $y$ are in a non-trivial component
		in $\sI_f$\,. Then, the \algorithmicfor\ loop of $\BuildCausalOrder$ will eventually
		encounter a vertex $v$ of which $x$ and $y$ are descendants.
		 
		In the depth-first traversal of influencing walks performed in $\TraverseInflWalk(G, I, O,
		f, \Sup, \status, v)$\,, eventually a directed cycle containing both $x$ and $y$ will be
		discovered. Without loss of generality, assume that the depth-first traversal starting from $v$
		visits $x$ before $y$\,: then, the depth-first traversal will eventually uncover a walk
		of the form
		\begin{align*}
				v
			\;\arc\;
				\cdots
			\;\arc\;
				x
			\;\arc\;
				\cdots
			\;\arc\;
				y
			\;\arc\;
				\cdots
			\;\arc\;
				y'
			\;\arc\;
				x	\;\;.
		\end{align*}
		Then in the procedure call $\TraverseInflWalk(G, I, O, f, \Sup, \status, y')$\,,
		line~\ref{line:recurAbortTest} will find $\status(x) = \pending$\,, as line~\ref{line:setPending}
		of the procedure call $\TraverseInflWalk(G, I, O, f, \Sup, \status, x)$ has been executed while
		line~\ref{line:fixStatus} has not. Then, the procedure aborts by returning $\mR$ without first
		changing the status of $\status(y')$ from $\pending$\,.
		 
		It is clear that if $w'$ depends on $w$\,, and if $\TraverseInflWalk(G, I, O, f, \Sup, \status, w')$
		aborts with $\status(w') = \pending$ during a procedure call $\TraverseInflWalk(G, I, O, f, \Sup, \status,
		w)$\,, then the latter will also abort with $\status(w) = \pending$\,. By induction, we may then show
		that for $v \in V(G)$ for which $x$ and $y$ are descendants, $\TraverseInflWalk(G, I, O, f, \Sup, \status, v)$
		will abort with $\status(v) = \pending$ in the \algorithmicfor\ loop in $\ComputeSuprema$\,.

		By the analysis of the case where $\preceq$ is antisymmetric, the status $\status(v) = \pending$
		will only occur at line~\ref{line:abortTest} of $\ComputeSuprema$ if $\preceq$ is not antisymmetric.
		If this occurs, $\sigma = \fail$\,; as well, no causal path cover exists for $(G,I,O)$ by
		Theorem~\ref{thm:uniqueFamilyPaths}, and thus no flow exists for $(G,I,O)$ by
		Theorem~\ref{thm:graphth-charn}. 
	\end{itemize}
	Thus, $\sigma \ne \fail$ iff $\preceq$ is a partial order; and when this occurs, by reduction to
	Algorithm~\ref{alg:fig3-8}, $\Sup$ corresponds to the natural pre-order $\preceq$ in the sense of
	Equation~\ref{eqn:correctPreorderRelation}.
\end{proof}

\paragraph{Run-time analysis.}

We may analyze the run-time of Algorithm~\ref{alg:computeSuprema} as follows.
\label{discn:runTimeBuildCausalOrder} Let $n = \abs{V(G)}$\,, $m = \abs{E(G)}$\,,
$k = \abs{I} = \abs{O}$\,, and $d$ be the maximum degree of $G$\,. The time required
to execute the \algorithmicforall\ loop starting on line~\ref{line:improveSuprema}
of \TraverseInflWalk\ is $O(k)$\,; then, aside from the work done in recursive invocations
to \TraverseInflWalk, the time required to perform an invocation of \TraverseInflWalk\ for
a vertex $v$ is $O(k \deg f(v))$\,.
Because the first invocation of \TraverseInflWalk\ for a vertex $v$ will change $\status(v)$ to
something other than \none, which prevents any further invocations for $v$\,, \TraverseInflWalk\ will
only be called once for any given vertex in the course of Algorithm~\ref{alg:computeSuprema}. Then,
summing over all vertices $v \in V(G)$\,, the amount of time required to perform the \algorithmicforall\
loop starting on line~\ref{line:beginTarjan} of \ComputeSuprema\ is $O(km)$\,. The time required by
$\InitStatus$ to initialize $\Sup$ and $\status$ is $O(kn)$\,; then, the overall running time of
Algorithm~\ref{alg:computeSuprema} is $O(km)$\,.

\subsubsection{A slightly more efficient algorithm for finding a causal order for $f$}

\setlength\unitlength{1em}
If $\mC$ is a causal path cover, it also is possible to find a causal order $\boxpreceq$
compatible with $f$ which differs from the natural pre-order for $f$, or determine that none exists,
by recursively assigning integer ``level'' values to vertices rather than building the set of
descendants. For example, one may construct a function $\lambda: V(G) \to \N$ satisfying
\begin{align}
		\begin{split}
			\lambda(x)
		\;=&\;\;
				0\,,
		\\
			\lambda(x)
		\;=&\;\;
				1 + \max \set{\lambda(y)}{x = f(y) \text{~~or~~} x \sim f(y)}\,,
		\end{split}
		&
		\begin{split}
			& \text{if there are no influencing walks for $\mC$ ending at $x$;}
		\\	
			 & \text{otherwise}.
		\end{split}
\end{align}
Note that the set $S(x)$ of vertices $y$ such that $x = f(y)$ or $x \sim f(y)$ are the initial points
for any influencing walk for $\mC$ with one segment which ends at $x$\,. By constructing the
predecessor function $g = f\inv$ of $\mC$ rather than the functions $P$ and $L$ in
Algorithm~\ref{alg:getChainDecomp}, we can easily find all elements of $S(x)$ in $G$ by
visiting $g(z)$ for $z = x$ or $z \sim x$\,. Then, such a level function can be constructed
by a Tarjan style algorithm similar to Algorithm~\ref{alg:traverseInflWalk}, using the $\status$
array in the same way, but traversing the arcs of the influencing digraph $\sI_f$ in the opposite
direction as $\TraverseInflWalk$\,. We may then define $x \;\boxpreceq\; y \;\;\iff\;\;
[x = y] \ou [\lambda(x) < \lambda(y)]$\,.

It is easy to see that the resulting partial order $\boxpreceq$ resulting would have the same
maximum-chain length as the natural pre-order $\preceq$\,: any maximal chain in $\preceq$ is a
list of the end-points of consecutive segments in an influencing walk for $\mC$\,, which will
be a maximum chain in $\boxpreceq$\,. However, $\boxpreceq$ also contains relationships between
vertices with no clear relation in the influencing digraph $\sI_f$\,, because it suffices for
two vertices to be on different ``levels'' for them to be comparable.

Such a causal order $\boxpreceq$ can actually be constructed in $O(m)$ time, because the algorithm
to construct it consists essentially of just a depth-first traversal with operations taking only
constant time being done at each step. We have instead presented the above algorithm because the
extra time required to obtain the coarsest compatible causal order for $f$ will not affect the
asymptotic run time of the complete algorithm for finding a flow, because of the immediate
reduction to the well-studied problem of transitive closure, and in the interest of describing
an algorithm to construct the natural pre-order for $f$ (being the coarsest compatible causal
order for $f$).

\subsection{The complete algorithm}
\label{sec:completeAlg}

We now describe the complete algorithm to produce a flow for a geometry $(G,I,O)$\,,
using Algorithms~\ref{alg:buildPathFamily} and~\ref{alg:computeSuprema}.
\begin{algorithm}[ht]
	\mycaption
		{\FindFlow(G,I,O)}
		{try to find a flow for $(G,I,O)$}
		\label{alg:findFlow}
	\begin{algorithmic}[1]
	\REQUIRE $(G,I,O)$ is a geometry with $\abs{I} = \abs{O}$

	\medskip

	\LET $\tau \gets \BuildPathFamily(G,I,O)$
			\label{line:buildPathFamily}
	\IFTHEN {$\tau = \fail$} \algorithmicreturn\ \fail

	\smallskip

	\LET $(f,P,L) \gets \GetChainDecomp(G,I,O,\tau)$
	\LET $\sigma \gets \ComputeSuprema(G,I,O,f,P,L)$
		\label{line:buildSuprema}
	\IF {$\sigma = \fail$}
		\RETURN \fail
	\ELSE
		\RETURN $(f, P, L, \sigma)$
	\ENDIF
	\end{algorithmic}
\end{algorithm}

\begin{corollary}
	Let $(G, I, O)$ be a geometry with $\abs{I} = \abs{O}$\,. Then $\FindFlow$ halts on input
	$(G,I,O)$\,. Furthermore, if $\FindFlow(G,I,O) = \fail$\,, then $(G,I,O)$ does not have a
	causal flow; otherwise, $\FindFlow(G,I,O) = (f, P, L, \Sup)$\,, and $(f, \preceq)$ is a causal
	flow, where $\preceq$ is characterized by
	\begin{align}
		\label{eqn:obtainedCausalOrder}
			x \preceq y
		\quad\iff\quad
			\Sup(P(y), x) \;\le\; L(y)	\,;
	\end{align}
	and is the natural pre-order for $f$\,.
\end{corollary}
\begin{proof}
	By Corollary~\ref{cor:buildPathCover}, a causal path cover exists for $(G,I,O)$ only if
	$\BuildPathFamily(G,I,O)$ sets $\tau \ne \fail$ on line~\ref{line:buildPathFamily}; thus if
	$\tau = \fail$\,, $(G,I,O)$ has no causal flow by Theorem~\ref{thm:graphth-charn}. Otherwise,
	$\tau$ is a path cover. If $\BuildCausalOrder$ sets $\sigma = \fail$ on line~\ref{line:buildSuprema},
	$(G,I,O)$ has no causal flow by Theorem~\ref{thm:buildCausalOrder}. Otherwise, the relation $\preceq$
	characterized by Equation~\ref{eqn:obtainedCausalOrder} is the natural pre-order for $f$ and a
	causal order, in which case $(f, \preceq)$ is a causal flow.
\end{proof}

\paragraph{Run-time analysis.}

Because $\tau \ne \fail$ at line~\ref{line:buildPathFamily} implies that $\tau$ is a path cover,
\GetChainDecomp\ visits each vertex $v \in V(G)$ once to assign values for $P(v)$\,, $L(v)$\,,
and possibly $f(v)$ in the case that $v \in O\comp$\,. Then, its' running time is $O(n)$\,.
The running time of \FindFlow\ is then dominated by \BuildPathCover\ and \ComputeSuprema,
each of which take time $O(km)$\,.

\section{Potential Improvements}
\label{sec:potentialImprovements}

This paper has described efficient algorithms for finding flows, with the aim of
not requiring prior knowledge of graph-theoretic algorithms in the presentation. This constraint has
led to choices in how to present the algorithms which may make them less efficient
(in practical terms) than may be achievable by the state of the art; and no significant
analysis of the graphs themselves have been performed. Here, I discuss
issues which may allow an improvement on the analysis of this article.

\subsection{Better algorithms for finding path covers}

For network-flow problems (the usual tools used for solving questions of maximum-size
collections of paths in graphs), there is a rich body of experimental results for efficient algorithms.
However, there seems to be very little discussion in the literature of the special
case where all edge capacities are equal to $1$\,, which is relevant to the problem
of finding maximum collections of vertex-disjoint $I$ -- $O$ paths. It is difficult
to determine, in this case, whether there is a significant difference in the performance
of various algorithms. Although it is less efficient than other algorithms for general
network flow problems, the most obvious choice of network flow algorithm for finding
a maximum family of vertex-disjoint $I$ -- $O$ paths is the Ford-Fulkerson algorithm,
which has an asymptotic running time $O(km)$\,. This running time is identical to
Algorithm~\ref{alg:buildPathFamily}: this should not be surprising, as 
Algorithm~\ref{alg:augmentSearch} essentially implements a depth-first variation
of the Ford-Fulkerson algorithm for finding an augmenting flow.

A more thorough investigation of network flows may yield an improved algorithm for
finding a path cover for $(G,I,O)$\,, which (when coupled with the faster algorithm
for finding a minimum-depth causal order) would yield a faster algorithm for finding
causal flows.

\subsection{Extremal results}

Consider all the ways we can add edges between $n$ vertices to get a geometry with
$k$ output vertices and a causal flow. Just to achieve a path cover, we require
$n - k$ edges; this lower bound is tight, as graph consisting of just $k$ vertex-disjoint
paths on $n$ vertices has this many edges, and the paths represent a causal path cover
of that graph. The more interesting question is of how many edges are required
to force a graph to not have any causal path covers. 

\def\p(#1,#2){p_{#1}^{\scriptscriptstyle (#2)}}

Let $\tilde{n}$ be the residue of $n$ modulo $k$\,. Consider a collection of $k$ paths
$\brace{P_j}_{j \in [k]}$\,, given by $P_j = \p(j,0) \cdots \p(j, \ceil{n/k} - 1)$
for $j < \tilde{n}$\,, and $P_j = \p(j, 0) \cdots \p(j, \floor{n/k} - 1)$ for $j \ge \tilde{n}$\,.
Then, let $G$ be the graph defined by adding the edges $\p(h, a) \p(j, a)$ for all $a$ and $h \ne j$ where
these vertices are well-defined, and $\p(j, a) \p(h, a+1)$ for all $a$ and $h < j$ where these vertices
are well-defined. We may identify the initial point of the paths $P_j$ as elements of $I$ and end-points
as elements of $O$\,: then, let $M(n,k)$ denote the geometry $(G,I,O)$ constructed in this way.

The geometry $M(n,k)$ has the obvious successor function given by $f(\p(j, a)) = \p(j, a+1)$
for all $j$ and $a$ where both vertices are defined. Then, consider the natural pre-order
for $f$\,:
\begin{romanum}
\item
	we obviously have $\p(j, a) \preceq \p(j, b)$ for $a \le b$\,, for every $j \in [k]$\,;

\item
	from the edges $\p(h, a) \, \p(j, a)$\,, we obtain $\p(h, a-1) \preceq \p(j, a)$ for
	all $h,j \in [k]$ and all $a > 0$\,; and

\item
	from the edges $\p(h ,b+1) \; \p(j, b)$ for $h < j$\,, we obtain $\p(h ,b) \preceq \p(j, b)$
	and $\p(j, b-1) \preceq \p(h, b+1)$\,. (Note that the second of these two constraints is redundant,
	as $\p(j,b-1) \preceq \p(h,b) \preceq \p(h,b+1)$ is implied by the above two cases.)
\end{romanum}
Then, the natural pre-order $\preceq$ on $M(n,k)$ is closely related to the lexicographical order
on ordered pairs: $\p(h,a)$ and $\p(j,b)$ are incomparable if they are both endpoints of their respective
paths $P_h$ and $P_j$\,, and otherwise $\p(h,a) \preceq \p(j, b)$ if and only if either $a < b$\,,
or $a = b$ and $h \le j$\,. This is clearly a partial order, so $M(n,k)$ has a causal flow: and it has
$kn - \binom{k+1}{2}$ edges in total.

I conjecture that this is the maximum number of edges that a geometry on $n$ vertices with $k$
output vertices can have. If this can be proven, we can determine that certain geometries have
no flows just by counting their edges; the upper bounds of this paper can then be improved to
$O(k^2 n)$\,.

\section{Open Problems}
\label{sec:openProbs}

To conclude, I re-iterate the open problems presented in~\cite{B06b}.

\begin{enumerate}
\item
	\textbf{The general case.}
	When $\abs{I} > \abs{O}$\,, it is easy to see that a causal flow cannot exist, because no
	successor function $f$ may be defined. This leaves the case where $\abs{I} < \abs{O}$\,. 
	If $\delta = \abs{O} - \abs{I}$\,, we may test sets $\partial I \subset I\comp$ with
	$\abs{\partial I} = \delta$ to see if the geometry $(G, I \union \partial I, O)$ has a causal
	flow: doing this yields an $O(kmn^\delta)$ algorithm for finding a causal flow for $(G,I,O)$\,. Is there an
	algorithm for finding causal flows in an arbitrary geometry with $\abs{I} \le \abs{O}$\,,
	whose run-time is also polynomial in $\delta = \abs{O} - \abs{I}$\,?

\item
	\textbf{Graphs without designated inputs/outputs.}
	Quantum computations in the one-way model may be performed by composing three
	patterns: one pattern to prepare an appropriate quantum state, a pattern to apply a unitary
	that state (in the vein that we have been considering in this article), and a final pattern 
	which measures the resulting state in an appropriate basis. The composite pattern has no
	input or output qubits, and so has only the measurement signals as an output. The result
	of the computation would then be determined from the parity of a subset of the measurement signals.

	Given a graph without any designated input or output vertices, what constraints
	are necessary to allow a structure similar to a causal flow to be found, which would 
	guarantee that deterministic $n$ qubit operations in the sense of~\cite{DK05} can be
	performed in the one-way measurement model with the entanglement graph $G$\,?

\item
	\textbf{Ruling out the presence of causal flows with only partial information about $G$\,.}
	Are there graphs $G$ where it is possible to rule out the presence of a flow for $(G,I,O)$
	from a proper sub-graph of $G$\,, or given $n = \abs{V(G)}$\,, $m = \abs{E(G)}$\,, and
	$k = \abs{I} = \abs{O}$\,?
	(This question obviously includes the extremal problem asked earlier.)

\item
	\textbf{Relaxing the causal flow conditions for Pauli measurements.}
	Suppose that, in addition to $I$ and $O$\,, we know which qubits are to be
	measured in the $X$ axis and which are to be measured in the $Y$ axis
	(corresponding to measurement angles $0$ and $\pi/2$ respectively). These
	qubits can always be measured first in a pattern, by absorbing byproduct
	operations on those qubits and performing signal shifting. However, the
	analysis of patterns in terms of causal flows does not take this into account, as
	it is independent of measurement angles. Is it possible to develop a natural
	analogue for causal flows which represents these qubits as
	minimal in the corresponding causal order, which may be efficiently found for
	geometries with $\abs{I} = \abs{O}$ or $\abs{I} \le \abs{O}$ generally?
\end{enumerate}

The results of this article were inspired by the similarity between of the characterization
in terms of causal flows, with aspects of graph theory related to Menger's Theorem in general,
and the relationship between influencing walks and alternating walks in particular. Investigation
into open questions involving efficient construction of causal flows or relaxations of them may
benefit from additional investigation of this link.

\paragraph{Acknowledgements.}

I would like to thank Elham Kashefi, who interested me in the problem of efficiently
finding causal flows for geometries and for helpful discussions; Donny Cheung and
Anne Broadbent, who provided useful insights; and Rob Raussendorf, for his feedback
on the presentation of this paper.

\section{Acknowledgements}

I would like to thank Elham Kashefi, who interested me in the problem of efficiently
finding causal flows for geometries and for helpful discussions; Donny Cheung, who
provided helpful remarks on sufficient conditions for a causal flow to exist; and
Anne Broadbent and Rob Raussendorf, for their contributions to my understanding of
the problem.

\bibliographystyle{amsalpha}

\end{document}

%% file: preamble-graphth-charn.tex
\usepackage{fullpage}
\usepackage{ifthen}
\usepackage{ifpdf}
\usepackage{amsthm}
\usepackage{amssymb}
\usepackage{amsmath}
\usepackage[mathscr]{euscript}
\usepackage{algorithm}
\usepackage{algorithmic}
\usepackage{ccaption}

\ifpdf
\usepackage[pdftex]{graphicx}
\else
\usepackage[dvips]{graphicx}
\fi

\usepackage{url}

\newcommand\ifthen[2] {\ifthenelse{#1}{#2}{}}
\newcommand\ifempty[3]{\ifthenelse{\equal{#1}{}}{#2}{#3}}

\newtheoremstyle{theorem}{5mm}{3mm}{\itshape}{}{\bfseries}{.}{1em}
		  {\thmname{#1}\thmnumber{ #2}\thmnote{ (#3)}}

\newtheoremstyle{theoremname}{5mm}{3mm}{\itshape}{}{\bfseries}{.}{1em}
		  {\thmname{\ifempty{#3}{#1}{#3}}}

\newtheoremstyle{dfn}{5mm}{3mm}{}{}{\bfseries}{.}{1em}{}

\newtheoremstyle{notation}{5mm}{3mm}{}{}{\bfseries}{.}{1em}{\thmname{#1}}

\newtheoremstyle{convention}{5mm}{3mm}{}{}{\itshape}{:}{1em}{\thmname{#1}}

\newtheoremstyle{proof}{0mm}{3mm}{}{}{\bfseries\itshape}{ ---}{1em}
		{\thmname{#1}\thmnumber{ #2}\thmnote{ #3}}

\theoremstyle{theorem}

\newtheorem{theorem}{Theorem}
\newtheorem{lemma}[theorem]{Lemma}

\newtheorem{corollary}[theorem]{Corollary}

\theoremstyle{theoremname}
\newtheorem*{theoremname}{(missing theorem name)}

\theoremstyle{dfn}

\newtheorem{definition}[theorem]{Definition}

\theoremstyle{notation}

\theoremstyle{convention}

\theoremstyle{proof}
  
\newtheorem*{pf}{Proof}

\renewenvironment{proof}[1][]
     {\ifempty{#1}{\begin{pf}}{\begin{pf}[#1]}}
     {\qed\end{pf}}

\let\amsmatrix\matrix       
\let\endamsmatrix\endmatrix 

\def \ge		{\geqslant}      
\def \le		{\leqslant}      
\renewcommand\preceq {\preccurlyeq}
\renewcommand\succeq {\succcurlyeq}
\def \x			{\times}         
\def \N			{\mathbb N}      

\def \implies			{\Longrightarrow}					
\def \iff				{\Longleftrightarrow}				
\def \inv				{^{-1}}          
\def \union				{\,\cup\,}       

\def \vide		{\varnothing}   
\def \ou		{\,\vee\,}      

\DeclareMathSymbol	{\inclnub}{\mathord}{letters}{44}

\DeclareMathSymbol	{\mapsnub}{\mathrel}{symbols}{55}

\def \to		{\longrightarrow}                                 
\def \smallto		{\rightarrow}					  
\def \mapsto		{\longmapsto}					  
\def \setminus		{\smallsetminus}				  
\DeclareMathSymbol	{\longsetminus}{\mathop}{symbols}{110}		  
\def \subset		{\subseteq}					  
\DeclareMathSymbol	{\prsubset}{\mathrel}{symbols}{26}		  
\DeclareMathSymbol	{\prsupset}{\mathrel}{symbols}{27}		  
\def \epsilon		{\varepsilon}					  
\DeclareMathSymbol	{\varphi}{\mathalpha}{letters}{39}		  
\DeclareMathSymbol	{\forallSymbol}{\mathord}{symbols}{56}		  
\DeclareMathSymbol	{\existsSymbol}{\mathord}{symbols}{57}		  

\newcommand  \boost[1]	{\big. #1 \big.}                       
\renewcommand\brace[1]	{\left\{ #1 \right\}}		       
\newcommand  \set[2]	{\brace{#1 \;\left|\; #2 \right.}}      
\newcommand  \floor[1]	{\left\lfloor #1 \right\rfloor}	       
\newcommand  \ceil[1]	{\left\lceil #1 \right\rceil}	       
\newcommand  \abs[1]	{\left| #1 \right|}		       
\newcommand  \ket[1]	{\boost{\left| #1 \right\rangle}\!}    

\renewcommand\bar[1]	{\overline{#1}}                        
\newcommand  \paren[1]	{\left( #1 \right)}		       
\newcommand  \sqparen[1]{\left[ #1 \right]}		       

\renewenvironment{matrix}
{\left[\:\! \amsmatrix}
{\endamsmatrix \:\!\right]}

\newenvironment{piecewise}
{\left\{ \begin{array}{c@{\quad}l} }
{\\ \end{array} \right\}}
\renewenvironment{cases}  
{\begin{piecewise}}
{\end{piecewise}}

\newenvironment{romanum}
{\vspace{-1.5ex}
\let\jrnbtemptopsep\topsep
\setlength{\topsep}{-1ex}
\begin{enumerate}
	\leftskip=1em
	
	\setlength{\itemsep}{-0.2ex}
}{\end{enumerate}\vspace{-1.5ex}
\setlength{\topsep}{\jrnbtemptopsep}}

\newcounter{alphanum}
	{\end{list}\vspace{-1.4ex}}

\newcounter{inlinum}
\renewcommand\theinlinum{\textbf{(\alph{inlinum})}}
\newenvironment{inlinum}{\renewcommand\item{\refstepcounter{inlinum}\theinlinum~\nolinebreak}}{}

\makeatletter
\def\iftwocol{\ifthenelse{\boolean{@twocolumn}}}
\makeatother

\newcommand\onecol[1]{\iftwocol{}{#1}}

\newcommand\arc\smallto

\newcommand\comp{^{\textsf{c}}}

\newcommand\mP{\mathcal{P}}
\newcommand\mC{\mathcal{C}}
\newcommand\mF{\mathcal{F}}

\newcommand\mR{\mathcal{R}}
\newcommand\sG{\mathscr G}
\newcommand\sI{\mathscr I}

\newcommand\sW{\mathscr W}
\newcommand\scrP{\mathscr{P}}

\newcommand\flowi{\textup{(\textsf{F}\textit{\!\!\;i})}}
\newcommand\flowii{\textup{(\textsf{F}\textit{\!\!\;i\!\!\;i})}}
\newcommand\flowiii{\textup{(\textsf{F}\textit{\!\!\;i\!\!\;i\!\!\;i})}}

\newcommand\algorithminput{\textbf{Input:}}
\newcommand\algorithmoutput{\textbf{Output:}}
\newlength\alginputlength
\newlength\alginputskiplength
\newcommand\fwdarc[3]{{#2 \textbf{~such that~} (#1 \arc #2) \in A(#3)}}

\newcommand\boxpreceq{%
	\setlength\unitlength{1em}%
	\begin{picture}(1,1)\put(0,-0.15){\framebox(1,1)[c]{$\preceq$}}\end{picture}}

\newcommand\typ[1]{\textup{\texttt{#1}}}

\newcommand\AugmentSearch{\typ{AugmentSearch}}
\newcommand\BuildPathCover{\typ{BuildPathCover}}
\newcommand\GetChainDecomp{\typ{GetChainDecomp}}
\newcommand\InitStatus{\typ{InitStatus}}
\newcommand\TraverseInflWalk{\typ{TraverseInflWalk}}
\newcommand\ComputeSuprema{\typ{ComputeSuprema}}

\newcommand\BuildPathFamily{\typ{BuildPathFamily}}

\newcommand\BuildCausalOrder{\typ{BuildCausalOrder}}
\newcommand\FindFlow{\typ{FindFlow}}

\newcommand\SimpleTC{\typ{SimpleTC}}
\newcommand\Root{\typ{Root}}
\newcommand\Comp{\typ{Comp}}
\newcommand\nil{\typ{nil}}
\newcommand\stack{\typ{stack}}
\newcommand\Desc{\typ{Desc}}
\newcommand\POP{\typ{POP}}
\newcommand\PUSH{\typ{PUSH}}

\newcommand\Sup{\typ{sup}}
\newcommand\status{\typ{status}}
\newcommand\fixed{\typ{fixed}}
\newcommand\pending{\typ{pending}}
\newcommand\none{\typ{none}}
\newcommand\fail{\typ{fail}}
\newcommand\success{\typ{success}}

\newcommand\iter{\typ{iter}}
\newcommand\visited{\typ{visited}}
\newcommand\prev{\typ{prev}}
\newcommand\next{\typ{next}}

\newcommand\AddArc{\typ{AddArc}}
\newcommand\RemoveArc{\typ{RemoveArc}}

\newlength{\commentskip}
\renewcommand\algorithmiccomment[1]{\\[-2.8ex] \hskip\commentskip \emph{\ldots\,#1} }
\renewcommand\algorithmicensure{\textbf{Result:}}
\renewcommand\algorithmicloop{\textbf{begin}}
\renewcommand\algorithmicendloop\algorithmicend
\newcommand\algorithmiclet{\textbf{let}}
\newcommand\algorithmicprocedure{\textbf{procedure}}
\newcommand\BEGIN\LOOP
\newcommand\END\ENDLOOP
\newcommand\IFTHEN[1]{\STATE \algorithmicif\ #1 \algorithmicthen\/}
\newcommand\PROCEDURE{\STATE \algorithmicprocedure\ }
\newcommand\LET{\STATE \algorithmiclet\ }
\newcommand\AND{\textbf{and}}

\newcommand{\captionfonts}{\footnotesize}

\makeatletter  
\long\def\@makecaption#1#2{%
  \vskip\abovecaptionskip
  \sbox\@tempboxa{{\captionfonts #1: #2}}%
  \ifdim \wd\@tempboxa >\hsize
    {\captionfonts #1: #2\par}
  \else
    \hbox to\hsize{\hfil\box\@tempboxa\hfil}%
  \fi
  \vskip\belowcaptionskip}
\makeatother   

\newcommand\mycaption[2]{\caption{\!\!\textbf{:}~~$#1$ --- #2}}
\newfixedcaption{\figcaption}{figure}
\def\puttext(#1,#2)[#3]#4{\put(#1,#2){\put(-0.5,0){\makebox(0,0)[#3]{#4}}}}

\newenvironment{citetheorem}[3][]%
	{\refstepcounter{theorem}\begin{theoremname}[#3~\thedefinition~{\protect \cite[#1]{#2}}]}%
	{\end{theoremname}}